\documentclass[journal]{IEEEtran}

\usepackage{cite}
\usepackage{empheq}
\usepackage{amsmath}
\usepackage{amsthm}
\usepackage{amssymb}
\usepackage{mathrsfs}
\usepackage{ulem}
\usepackage{graphicx}
\usepackage{algorithm}
\usepackage{enumerate}
\usepackage{url}
\graphicspath{{Figs/}{../pdf/}{../jpeg/}}
\DeclareGraphicsExtensions{.pdf,.jpeg,.png}
\usepackage{cancel}
\usepackage{makecell}
\usepackage{bbding}
\usepackage{booktabs}
\usepackage{array}
\usepackage[hidelinks]{hyperref}

\usepackage{xcolor}
\definecolor{myblue}{RGB}{0,112,192}

\begin{document}

\title{A Data-Driven Optimal Control Architecture for Grid-Connected Power Converters}

\author{Ruohan~Leng, Linbin~Huang, Huanhai~Xin, Ping~Ju, Xiongfei~Wang, \\
Eduardo~Prieto-Araujo, and~Florian~D{\"o}rfler
\vspace{-7mm}
\thanks{This work was supported by the National Natural Science Foundation of China under Grant U24B6008.}
\thanks{Ruohan Leng, Linbin Huang, Huanhai Xin, and Ping Ju are with the College of Electrical Engineering, Zhejiang University, Hangzhou 310027, China. (e-mail: {\{lengruohan, hlinbin, xinhh, pju\}}@zju.edu.cn).}
\thanks{Xiongfei Wang is with the Department of Electrical Engineering, Tsinghua University, Beijing, China. (e-mail: xiongfei@tsinghua.edu.cn).}
\thanks{Eduardo Prieto-Araujo is with Universitat Politècnica de Catalunya (UPC), Barcelona, Spain. (e-mail: eduardo.prieto-araujo@upc.edu).}
\thanks{Florian D{\"o}rfler is with the Department of Information Technology and Electrical Engineering at ETH Z{\"u}rich, Switzerland. (e-mail: dorfler@ethz.ch).}
}
	\maketitle
    
	\begin{abstract}
	Grid-connected power converters are ubiquitous in modern power systems, acting as grid interfaces of renewable energy sources, energy storage systems, electric vehicles, high-voltage DC systems, etc. Conventionally, power converters use multiple PID regulators to achieve different control objectives such as grid synchronization and voltage/power regulation, where the PID parameters are usually tuned based on a presumed (and often overly-simplified) power grid model. However, this may lead to inferior performance or even instabilities in practice, as the real power grid is highly complex, variable, and generally unknown. To tackle this problem, we employ a data-enabled predictive control (DeePC) to perform data-driven, optimal, robust, and adaptive control for power converters. We call the converters that are operated in this way \textit{DeePConverters}. A DeePConverter can implicitly perceive the characteristics of the power grid from measured data and adjust its control strategy to achieve optimal, robust, and adaptive performance. We present the modular configurations, generalized structure, control behavior specification, inherent robustness, detailed implementation, computational aspects, and online adaptation of DeePConverters. High-fidelity simulations and hardware-in-the-loop (HIL) tests are provided to validate the effectiveness of DeePConverters.

	\end{abstract}
	
	
	\section{Introduction}\label{sec:intro}
	
	To tackle climate change, countries from all over the world have made ambitious plans to integrate renewable energy sources into their power grids. For instance, China reached a renewable energy (wind/solar) share of over 40\% in its total installed power generation capacity by the end of 2024~\cite{nea2025report}, while the EU aims to achieve 45\% by 2030~\cite{plan2018communication}.
    Unlike conventional fossil-fuel-based power plants that rely on synchronous generators (SGs) for power transfer and grid connection, renewable energy sources are connected to the power grid via {power electronics converters}~\cite{milano2018foundations, rocabert2012control, wang2020grid}.
	
	Compared with SGs, power converters have low physical inertia, suffer from the intermittency of renewables, and are distributed. Hence, their large-scale integration dramatically changes the characteristics of power systems and poses challenges to the stability analysis and secure operation of modern power grids. On the other hand, converters have fast actuation and high flexibility in controlling the voltage angle and magnitude, which, given appropriate control design, can potentially ensure a resilient power electronics-dominated power system. However, there exist a lot of challenges in designing reliable controllers for grid-connected power converters due to the strong interaction between the converter and the grid~\cite{sun2011impedance, wang2017unified}.
	
	Conventionally, a power converter employs multiple nested PID-based control loops to achieve multiple control objectives, e.g., grid synchronization, current control, and power/voltage regulation~\cite{blaabjerg2006overview, d2015virtual}. The control structure (i.e., how different loops are connected) is usually obtained based on engineering experience, and the corresponding PID parameters are tuned manually using simulations/experiments of a single-converter-infinite-bus system in a fit-and-forget fashion. However, the real power grid, which interacts with the converter in closed loop, has complex and time-varying dynamics and cannot be approximated by an infinite bus, especially in the context of power electronics-dominated power systems. This \textit{model mismatch} (between the real-world power grid model and the presumed model for converter design) may degrade the control performance and even result in instabilities. For instance, oscillations (small-signal instabilities) have been observed in many real-world wind farms~\cite{liu2017subsynchronous, cheng2020}. Before connecting to the grid, the wind turbines were thoroughly tested when they are connected to an ideal voltage source (infinite bus), yet instabilities can still arise in the real world due to the model mismatch. This problem can potentially be resolved by taking into account the model of the real power grid in the control design. Nonetheless, the real power grid is variable, high-dimensional, subject to lots of uncertainties, and thus generally unknown to the control designer of power converters. 
  
    A natural remedy is to employ advanced non-data-driven control strategies, such as robust control (e.g., $H_\infty$ control~\cite{huang2020h} or $\mu$-synthesis~\cite{hakemi2021generic}), adaptive control~\cite{el2021adaptive}, and model predictive control (MPC)~\cite{kouvaritakis2016model}, to cope with the uncertainties and improve stability margins. These approaches can deliver satisfactory performance when a sufficiently accurate converter-grid model and reliable retuning strategies are available. However, in power electronics-dominated power systems, such prerequisites are often not guaranteed due to the uncertain and time-varying grid impedance, complex converter-grid interactions, and proprietary device details. Hence, maintaining an accurate model and repeatedly retuning the controller can be costly and may lag behind the time-varying system dynamics.

	In fact, since a power converter interacts with the connected power grid in closed loop, it should be able to ``perceive'' the dynamics of the power grid during the on-line operation. According to the behavioral systems theory, the dynamics of a linear time-invariant (LTI) system can be fully captured by its trajectories (or equivalently, input/output data) given that the system is controllable and the input data is persistently exciting~\cite{willems2005note,markovskydata}. This result, also known as the \textit{Fundamental Lemma}, was formulated by Willems et al. and has received renewed attention recently as it allows a data-centric representation of an LTI system. For a grid-connected power converter, the Fundamental Lemma indicates that the closed-loop (i.e., converter-grid) dynamics are hidden in the current/voltage/power data. Now the question becomes: what can we do with the data if it captures the system dynamics?
	Recently, multiple direct data-driven control approaches have been proposed based on the Fundamental Lemma, e.g.~\cite{coulson2019data, de2019formulas, berberich2019data, huang2023robust}, which shows that one can obtain optimal control policies from data without identifying a model of the system. Since data can be conveniently obtained/recorded in the controller (usually a digital signal processor) of a converter, it is possible to use data to derive an optimal and robust controller that takes into account the unknown power grid dynamics and thus handles the model mismatch problem. 

	In our previous works \cite{huang2019data} and~\cite{huang2021decentralized}, we applied a novel data-enabled predictive control (DeePC) as a data-driven auxiliary control loop in power converters to eliminate power system oscillations, which shows the potential of using data to ``learn'' the system dynamics and design stabilizing controllers. DeePC only requires input/output data to predict future behaviors and compute robust and optimal control inputs for the system. DeePC has also been successfully applied in many neighboring engineering domains~\cite{carlet2020data, lian2023adaptive, elokda2021data, fawcett2022toward}. Our previous work~\cite{huang2021quadratic} applied DeePC to grid-connected converters by replacing certain control loops (specifically, the power regulation and synchronization layers) with a data-driven controller, which demonstrates fast tracking performance in both simulations and experiments. However, this line of work has barely scratched the surface: it has yet to explore full replacement of the control structure with DeePC to coordinate other control objectives, generalization across existing schemes like grid-following (GFL) and grid-forming (GFM) controls, and further improvement of grid-friendly behaviors. Moreover, the problem of steady-state errors caused by nonlinearity has not been addressed so far.
	
	In this paper, we apply DeePC to comprehensively perform data-driven, optimal, and robust control of voltage source converters (VSCs), focusing on how to replace conventional nonadaptive control loops to enhance power system performance and ensure stability. We refer to converters operated in this manner as \textit{DeePConverters}. As a first step, we introduce the modular implementations and an integral form for DeePConverters to respectively enable flexible configurations of their functionalities and eliminate tracking errors caused by system nonlinearities. Moreover, we demonstrate how a generalized structure of DeePConverters includes both GFL and GFM control schemes and discuss the specification of the DeePConverter's behaviors, such as voltage/power regulation and synchronization, as well as the realization of GFL and GFM operations via cost function design. We demonstrate the inherent robustness of DeePConverters to data disturbances, highlighting their robustness in realistic power system conditions. We further present efficient real-time implementations using high-performance solvers and closed-form solutions, and validate the benefits of DeePConverters through high-fidelity simulations and hardware-in-the-loop (HIL) tests in both converter-level and system-level settings. Finally, we explore how DeePConverters can achieve adaptability, enabling self-adjustments to time-varying dynamics and ensuring reliable operation over time. This data-driven paradigm synthesizes control directly from short measured input/output trajectories, thereby mitigating model mismatch and reducing reliance on explicit models and repeated retuning. The contributions of this paper can be summarized as follows:

     1) Based on the behavioral “data-as-model” paradigm, we propose a modular plug-and-play architecture to enable the data-driven optimal control of grid-connected power converters. This architecture allows flexible and selective replacements of the key control layers (e.g., synchronization and power/voltage regulation) or even the entire control scheme, facilitating the practical implementation of DeePC in power electronics converters.

    2) We employ an integral and adaptive DeePC formulation within the DeePConverter framework by incorporating discrete-time integral actions and allowing online updates of the Hankel matrices. The integral action eliminates steady-state tracking errors, while the recursive update and batch reconstruction strategies enhance the adaptability to time-varying conditions commonly encountered in converters.

    3) We investigate the cost function design in a predictive control setting to systematically realize and tailor the grid synchronization and power/voltage regulation functionalities in converters. Through structured design of the weighting and coupling matrices, our approach can emulate existing control paradigms (GFL/GFM) and configure the operating modes (e.g., PQ/PV), enabling direct and data-driven synthesis of control behaviors.

    The rest of this paper is organized as follows: Section II gives a brief review of DeePC and introduces the integral DeePC. Section III presents the modular implementations and an integral form for DeePConverters. Section IV discusses the generalized structure, control behavior specification, inherent robustness, implementation, computation, and online adaptation of the DeePConverters. Section V presents high-fidelity simulations and HIL tests. Section VI concludes the paper.

    \textit{Notation:} Let $\mathbb Z$ denote the set of integers, and $[n] = \{ 1, ..., n \}$ the index set with cardinality $n\in \mathbb Z_{>0}$. For a vector $x$, we use $\|x\|_2$ to denote its 2-norm, and use $\| x \|_A^2$ to denote  $x^\top A x$. We use $\operatorname{tr}(\cdot)$ to denote the trace of a square matrix. We use $\boldsymbol 1_n$ to denote a vector of ones of length $n$, and $I_n$ to denote an $n$-by-$n$ identity matrix (abbreviated as $I$ when the dimensions can be inferred from the context). We use $A^+$ to denote the pseudoinverse of the matrix $A$. We use \(\Pi\) to denote an orthogonal projection operator. We use $[Z_0;Z_1;...;Z_\ell]$ to denote the matrix $[Z_0^{\top}\; Z_1^{\top}\; \cdots \; Z_\ell^{\top}]^{\top}$. We use $\otimes$ to denote the Kronecker product. We use $\mathcal{D}_n$ to denote the first-order difference operator of size $n$.
	
	\section{Data-Enabled Predictive Control}

    In this section, we give a brief review of DeePC, and then introduce the integral DeePC to mitigate tracking errors caused by nonlinearities in converter systems. Though many components in converters and power grids can be modeled as continuous-time systems (e.g., filters and SGs), the controllers are generally in discrete time.
    Hence, we discretize the continuous-time components and consider the overall system as a discrete-time system (with a sufficiently small periodic sampling time to accurately capture system dynamics).
	
	\subsection{The role of input/output data}
	
    We first show how the system dynamics can be represented by input/output data.
	Consider a discrete-time LTI system
	\begin{equation}
		\left\{ \begin{array}{l}
			{x_{t + 1}} = A{x_t} + B{u_t}\\
			{y_t} = C{x_t} + D{u_t},
		\end{array} \right.\,		\label{eq:ABCD}
	\end{equation}
	where $x_t \in \mathbb{R}^n$ is the state of the system at~$t \in \mathbb{Z}_{ \ge 0}$,~$u_{t} \in \mathbb{R}^m$ is the input vector, and $y_{t} \in \mathbb{R}^p$ is the output vector. We assume that the system \eqref{eq:ABCD} is a minimal realization, meaning it is both controllable and observable. We collect length-$T$ input and output trajectories (i.e., from time $0$ to time $T$ where $T \in \mathbb{Z}_{ \ge 0}$) from the system, denoted by
\begin{equation*}
\begin{split}
   u^{\rm d} = \; & [u_0;u_1;\dots; u_{T-1}]\in \mathbb{R}^{mT} ,\\
   y^{\rm d} = \; & [y_0;y_1; \dots; y_{T-1}]\in \mathbb{R}^{pT} ,
\end{split}
\end{equation*}
which are data sequences that will be used to capture the system dynamics.
We use the input trajectory $u^{\rm d}$ to construct the Hankel matrix of depth $L$ as
\begin{equation}
		\mathscr{H}_L(u^{\rm d}) := {\small \begin{bmatrix}
				{{u_0}}&{{u_1}}& \cdots &{{u_{T - L}}}\\
				{{u_1}}&{{u_2}}& \cdots &{{u_{T - L + 1}}}\\
				\vdots & \vdots & \ddots & \vdots \\
				{{u_{L-1}}}&{{u_{L}}}& \cdots &{{u_{T-1}}}
		\end{bmatrix}}  \,.		
		\label{eq:Hankel_L}
\end{equation}
Similarly, for the output trajectory $y^{\rm d}$, we construct the Hankel matrix $\mathscr{H}_L(y^{\rm d})$. Then, we partition the Hankel matrices into
	\begin{equation*}
		[U_{\rm P};U_{\rm F}] := \mathscr{H_c}_{T_{\rm ini}+N}(u^{\rm{d}}) \quad \text{and} \quad [Y_{\rm P};Y_{\rm F}] := \mathscr{H_c}_{T_{\rm ini}+N}(y^{\rm{d}})\,,		\label{eq:partition_Huy}
	\end{equation*}
where $U_{\rm P} \in \mathbb{R}^{mT_{\rm ini} \times H_c}$, $U_{\rm F} \in \mathbb{R}^{mN \times H_c}$, $Y_{\rm P} \in \mathbb{R}^{pT_{\rm ini} \times H_c}$, $Y_{\rm F} \in \mathbb{R}^{pN \times H_c}$, and $H_c = T-T_{\rm ini}-N+1$. Here the subscript P (for ``past'') indicates that the matrices will be used to implicitly estimate the system's initial condition, and the subscript F indicates that the matrices will be used to predict the ``future'' behavior of the system.

The input and output trajectories should be sufficiently long and ``rich'' to fully capture the system dynamics. The so-called Fundamental Lemma in~\cite{willems2005note} gives explicit sufficient conditions on how long and ``rich'' the trajectories should be, namely, 1) the length of the trajectories should satisfy $T \ge (m+1)(T_{\rm ini} + N + n) - 1$, and 2) the input trajectory $u^{\rm d}$ should be persistently exciting of order $T_{\rm ini} + N + n$, i.e., $\mathscr{H_c}_{T_{\rm ini}+N+n}(u^{\rm{d}})$ should be of full row rank.
Under such conditions, the image of $\mathscr{H}_{T_{\rm ini}+N}(u^{\rm{d}},y^{\rm{d}})$ spans all length-$(T_{\rm ini}+N)$ trajectories, indicating that $[u_{\rm ini};u;y_{\rm ini};y]$ is a trajectory of (\ref{eq:ABCD}) if and only if there exists $g \in \mathbb{R}^{H_c}$ so that
\begin{equation}\label{eq:Hankel_g}
[
	{{U_{\rm P}}};
	{{Y_{\rm P}}};
	{{U_{\rm F}}};
	{{Y_{\rm F}}}
	]g = [
	{{u_{\rm ini}}};
	{{y_{\rm ini}}};
	u;
	y
	]\,,		
\end{equation}
where $u_{\rm ini} \in \mathbb R^{mT_{\rm ini}}$, $y_{\rm ini} \in \mathbb R^{pT_{\rm ini}}$, $u \in \mathbb R^{mN}$, $y \in \mathbb R^{pN}$.
The vector $[u_{\rm ini};y_{\rm ini}] $ can be thought of as the initial trajectory where $[u;y]$ departs.
If $T_{\rm ini}$ is larger than the lag of the system~\eqref{eq:ABCD} (i.e., the smallest integer $\ell \in \mathbb{Z}_{\ge 0}$ such that the observability matrix $[C;CA;...;CA^{\ell-1}]$ has rank $n$), for every given future input $u$, the future output $y$ is uniquely determined through (\ref{eq:Hankel_g}) \cite{markovsky2008data}. A necessary and sufficient condition was further derived in~\cite{markovskyidentifiability}, which only requires rank$\left(\mathscr{H}_{T_{\rm ini}+N}(u^{\rm{d}},y^{\rm{d}})\right)=m(T_{\rm ini}+N) + n$. This condition extends and includes the original Fundamental Lemma, and allows us to use other matrix structures beyond Hankel matrices as data-driven predictors.

\subsection{Review of DeePC}
	
	Based on~\eqref{eq:Hankel_g}, the DeePC method proposed in \cite{coulson2019data} directly uses input/output data collected from the unknown system to perform safe and optimal control without identifying a model. Moreover, robust and regularized DeePC solves the following optimization problem to obtain the optimal future inputs~\cite{huang2023robust}
	\begin{equation} 			\label{eq:DeePC} %\tag{DeePC}
		\begin{array}{l}
			\displaystyle \mathop {{\rm{min}}}\limits_{g, \sigma_{u}, \sigma_{y} \atop [u; y] \in \mathcal C } \;  {\left\| u \right\|_R^2} + {\left\| {y - r} \right\|_Q^2} + \lambda_{u} \| \sigma_{u} \|_2^2 + \lambda_{y} \| \sigma_{y} \|_2^2 + \lambda_g h(g) \\ \;\;\;\; {\rm s.t.} \;\;\; [
	U_{\rm P};
	Y_{\rm P};
	U_{\rm F};
	Y_{\rm F}
	]g = [
	u_{\rm ini} + \sigma_{u};
	y_{\rm ini} + \sigma_{y};
	u;
	y] ,
		\end{array}
	\end{equation}
	where $\mathcal C$ is the set of input/output constraints, $\sigma_u$ and $\sigma_y$ are slack variables to soften the equality constraints, which are penalized in the cost function with nonnegative coefficients $\lambda_u$ and $\lambda_y$. A regularization term \( h(g) \) is added with a tunable coefficient \( \lambda_g \). There are a few options~\cite{markovskydata}, and a typical choice is \( h(g) = \|g\|_2^2 \). The positive definite matrix~$R$ and positive semi-definite matrix~$Q$ are the cost matrices. The vector $r \in \mathbb{R}^{pN}$ is a reference trajectory for the outputs.

    Note that the dynamics of the to-be-controlled system are captured by the equality constraints in~\eqref{eq:DeePC}.
    We refer to~\cite{markovskydata} for a detailed discussion and review on the formulation~\eqref{eq:DeePC}.
    Similar to MPC, DeePC solves \eqref{eq:DeePC} in a receding horizon fashion, that is, after calculating the optimal control sequence $u^\star =: [u_0^\star;u_1^\star; \dots ; u_{k-1}^\star]$, we apply $(u_t,...,u_{t+k-1}) = (u_0^{\star},...,u_{k-1}^{\star})$ to the system for $k \le N$ time steps (i.e., the control horizon is $k$), then, reinitialize~\eqref{eq:DeePC} by updating $[u_{\rm ini};y_{\rm ini}]$ to the most recent measured trajectory, and setting $t$ to $t+k$, to calculate the new control sequence for the next $k$ time steps.

\subsection{Integral DeePC}\label{integral1}

Though the Fundamental Lemma applies to LTI systems, control engineers across different domains often observe satisfactory performance when applying DeePC to weakly nonlinear systems. However, we also observe that sometimes the nonlinearity may result in tracking errors. To alleviate this issue, we introduce an integral action into the DeePC framework to eliminate steady-state tracking errors, functioning similarly to the integrator in PID controllers but derived from a data-driven perspective.
As a first step, we reformulate \eqref{eq:ABCD} as
\begin{equation}
			\begin{bmatrix} \Delta x_{t+1} \\ y_t \end{bmatrix} = \begin{bmatrix}  A & 0 \\ C & I \end{bmatrix} \begin{bmatrix} \Delta x_{t} \\ y_{t-1} \end{bmatrix} + \begin{bmatrix} B \\ D \end{bmatrix} \Delta u_t \,, \label{eq:ABCD_Delta}
\end{equation}
where $\Delta x_t = x_t - x_{t-1}$ and $\Delta u_t = u_t - u_{t-1}$. In \eqref{eq:ABCD_Delta}, the (augmented) state variable is $[ \Delta x_t ; y_{t-1} ]$, the control input is $\Delta u_t$, while the output of the system is still $y_t$. Note that \eqref{eq:ABCD_Delta} is equivalent to \eqref{eq:ABCD}, but it allows us to focus on how the difference of the original control input (i.e., $\Delta u_t$) affects the system.
To capture the dynamics of system~\eqref{eq:ABCD_Delta}, we collect
\begin{equation*}
\Delta u^{\rm d} = [\Delta u_0;\Delta u_1;\dots; \Delta u_{T-1}]\in \mathbb{R}^{mT} \,,
\end{equation*}
and construct the Hankel matrix
\begin{equation*}
[\Delta U_{\rm P};\Delta U_{\rm F}] := \mathscr{H_c}_{T_{\rm ini}+N}(\Delta u^{\rm{d}}) \,.
\end{equation*}

Since \eqref{eq:ABCD_Delta} is still LTI, similar to~\eqref{eq:Hankel_g}, $[\Delta u_{\rm ini};\Delta u;y_{\rm ini};y]$ is a trajectory of \eqref{eq:ABCD_Delta} if and only if there exists $g \in \mathbb{R}^{H_c}$ so that
\begin{equation}\label{eq:Hankel_g_Delta}
[
	{{\Delta U_{\rm P}}};
	{{Y_{\rm P}}};
	{{\Delta U_{\rm F}}};
	{{Y_{\rm F}}}
	]g = [
	{{\Delta u_{\rm ini}}};
	{{y_{\rm ini}}};
	\Delta u;
	y
	]\,.		
\end{equation}

Then, we consider the following integral DeePC problem
\begin{equation}\label{eq:DeePC_integral}
%\tag{Integral-DeePC}
		\begin{array}{l}
		\hspace{-3mm}\displaystyle \mathop {{\rm{min}}}\limits_{g, \sigma_{u}, \sigma_{y} \atop [\Delta u; y] \in \mathcal C } \;
 {\left\| \Delta u \right\|_{R_\Delta}^2} + {\left\| {y - r} \right\|_Q^2} 
  + \lambda_{u} \| \sigma_{u} \|_2^2 + \lambda_{y} \| \sigma_{y} \|_2^2 + \lambda_g h(g)  \\
\;\;\;\; \hspace{-3mm}{\rm s.t.} \;\;\; [
	\Delta U_{\rm P};
	Y_{\rm P};
	\Delta U_{\rm F};
	Y_{\rm F}
	]g  = [
	\Delta u_{\rm ini} + \sigma_{u};
	y_{\rm ini} + \sigma_{y};
	\Delta u;
	y] , \vspace{2mm} \\
\hspace{11.7mm}   u = {\bf 1}_{N} \otimes I_m u_{t-1} + \underbrace{ \begin{bmatrix}
1&0& \cdots &0\\
1&1& \cdots &0\\
\vdots & \vdots & \ddots & \vdots \\
1&1& \cdots &1
		\end{bmatrix}}_{\in \mathbb{R}^{N \times N}} \otimes I_m \Delta u ,
		\end{array}
\end{equation}
%{\left\| u \right\|_R^2}
where $u_{t-1}$ denotes the (recorded) control input of system \eqref{eq:ABCD} at time $t-1$ when solving~\eqref{eq:DeePC_integral} at time $t$. We penalize the control input difference $\Delta u$ in the cost function with a positive definite cost matrix $R_\Delta$.
Similar to DeePC, we implement \eqref{eq:DeePC_integral} in a receding horizon manner. The difference is that the integral DeePC focuses on the input difference $\Delta u$. For instance, consider a control horizon of $k=1$ and let $\Delta u^\star (t) =: [\Delta u_0^\star(t); \Delta u_1^\star(t); \dots; \Delta u_{N-1}^\star(t)]$ be the optimizer of~\eqref{eq:DeePC_integral} at time $t$; we apply $u_t = u_{t-1} + \Delta  u_0^\star(t)$ to~\eqref{eq:ABCD} at time $t$, and apply $u_{t+i} = u_{t-1} + \sum\nolimits_{j = t}^{t+i} \Delta  u_0^\star(j)$ at time $t+i$, which is equivalent to adding a (discrete-time) integrator between $\Delta u_0^\star (t)$ and the system~\eqref{eq:ABCD}. Since $\Delta u_0^\star (t)$ is implicitly a function of the output tracking error according to~\eqref{eq:DeePC_integral}, this integral action has similar effects to the integrator in PID control and thus helps eliminate tracking errors. We note that integral actions are also often used in MPC to improve robustness and achieve perfect tracking~\cite{Verheijen2023HandbookDPC,lovaas2010robust}.

\section{Modular Implementations and Integral Form for DeePConverters}
Although the physical power grid is complex and unknown from the perspective of power converters, its dynamics can be implicitly captured via input/output data.
Inspired by this, we introduce a set of modular DeePConverter implementations by replacing the VSC's conventional nonadaptive control loops with a DeePC controller, enabling flexible data-driven control while maintaining compatibility with standard schemes. In addition, an integral form for DeePConverters is presented to improve the steady-state tracking accuracy.

\subsection{VSC System Descriptions}

\begin{figure}[!t]
 \vspace{0mm}
\centerline{\includegraphics[width=0.94\linewidth]{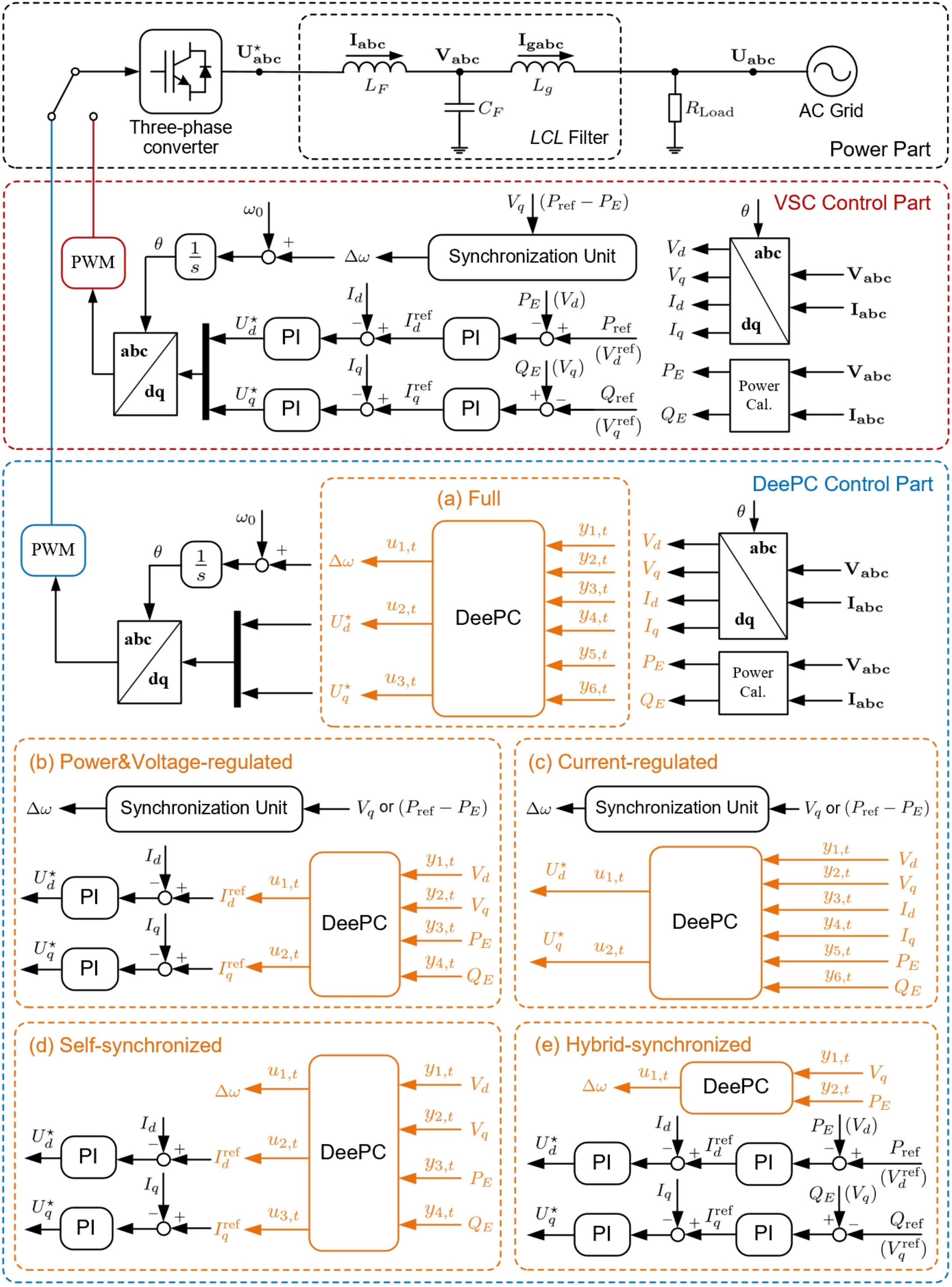}}
	\vspace{-2mm}
        \caption{One-line diagram of a three-phase power converter with conventional VSC control scheme and five optional DeePConverter configurations. We use $u_{t,i}$ ($i \in [m]$) to denote the $i$-th element of the input $u_t$ and $y_{t,i}$ ($i \in [p]$) to denote the $i$-th element of the output $y_t$.}
        \label{Fig_converter_1}
	\vspace{-2mm}
\end{figure}

Fig.~\ref{Fig_converter_1} shows a three-phase power converter connected to the grid via an \textit{LCL} filter. The three-phase voltage and current are represented by two-dimensional vectors in the synchronously rotating $dq$ frame via Park transformation. As labeled in Fig.~\ref{Fig_converter_1}, $U^*_{dq}$ denotes the converter-side voltage vector generated by PWM; $I_{dq}$ and $I_{gdq}$ are the converter-side and grid-side current vectors, respectively; $V_{dq}$ is the capacitor voltage of the \textit{LCL} filter; and $U_{dq}$ is the grid-side voltage vector. 

A VSC control scheme typically comprises three components: a synchronization unit, an outer loop, and an inner current loop~\cite{rocabert2012control}, as shown in the red box of Fig.~\ref{Fig_converter_1}. The synchronization unit generates the $dq$ frame frequency $\Delta \omega + \omega_0$, where $\omega_0$ is the nominal frequency. In general, two synchronization methods are commonly adopted. When the converter relies on a phase-locked loop (PLL) to align with the grid voltage~\cite{guo2023inertial}, it operates in GFL mode. In contrast, when the converter establishes its frequency reference through active power feedback by emulating a swing equation, e.g., a virtual synchronous machine (VSM), or via droop control~\cite{huang2017transient}, it operates in GFM mode. The outer loop computes the current references: $I^{\mathrm{ref}}_d$ is typically derived from active power control or $d$-axis voltage control, and $I^{\mathrm{ref}}_q$ from reactive power control or $q$-axis voltage control. The inner current loop, implemented in the $dq$ frame, ensures fast tracking of the current references.

In this paper, the considered GFL converters adopt PLL-based synchronization and PQ outer loops, whereas the GFM converters adopt swing-equation-based synchronization and AC voltage outer loops. The corresponding control structures of the GFL and GFM converters are the same as those in~\cite{yang2020placing}.

\subsection{Modular Implementations for DeePConverters}

Under the control structure of VSCs, which includes a synchronization unit and cascaded outer and current loops, there are five DeePC configuration options that can be used based on different combinations of control components being replaced, as shown in Fig.~\ref{Fig_converter_1}. These configurations are:

\textit{a) Full DeePConverter}: The synchronization unit, outer and current loops are all replaced by DeePC, as shown in Fig.~\ref{Fig_converter_1}~(a);

\textit{b) Power\&Voltage-regulated DeePConverter}: Only the outer loop is replaced by DeePC, as shown in Fig.~\ref{Fig_converter_1}~(b);

\textit{c) Current-regulated DeePConverter}: Both outer and current loops are replaced by DeePC, as shown in Fig.~\ref{Fig_converter_1}~(c);

\textit{d) Self-synchronized DeePConverter}: Synchronization unit and outer loop are replaced by DeePC, as shown in Fig.~\ref{Fig_converter_1}~(d);

\textit{e) Hybrid-synchronized DeePConverter}: Only the synchronization unit is replaced by DeePC, as shown in Fig.~\ref{Fig_converter_1}~(e).

These modular configurations enable flexible data-driven control compatible with conventional VSC schemes, allowing selective replacement of specific control layers based on application needs. This modular design is particularly valuable in industrial applications, where certain modules, such as synchronization units or current controllers, have already been pre-certified or fixed during the development phase. In particular, configurations (b) and (c) both involve selective replacement of the outer loop, which aim to achieve optimal power tracking or voltage regulation performance, while preserving the original synchronization unit. Configuration (e) replaces the synchronization unit, aiming to achieve an optimally engineered hybrid synchronization mechanism driven by a tunable combination of $q$-axis voltage and active power. In contrast, configurations (a) and (d) replace both synchronization unit and outer loop, leading to a more generalized control structure that can conveniently realize either GFL or GFM operation.

\subsection{Integral DeePConverters for Steady-State Accuracy}

Among the configurations in Fig.~\ref{Fig_converter_1}, certain designs, such as the Full and Power\&Voltage-regulated DeePConverters, directly perform power/voltage regulation. In these cases, accurate steady-state tracking becomes particularly crucial, as even small steady-state errors can accumulate over time, leading to long-term performance degradation or system imbalance.

To further enhance the steady-state regulation capability, the DeePConverter formulation can be extended by incorporating an integral structure, as previously discussed in subsection~\ref{integral1}. Specifically, integrators are inserted into the control paths so that DeePC optimizes the control input differences instead of the inputs themselves. The actual control inputs are then recovered by summing up these differences over time. This integral mechanism enables the controller to retain memory of past tracking errors and apply persistent corrections, thereby improving long-term tracking~\cite{lovaas2010robust}, especially in applications requiring accurate power regulation.

\begin{figure}[!t]
 \vspace{0mm}
\centerline{\includegraphics[width=0.8\linewidth]{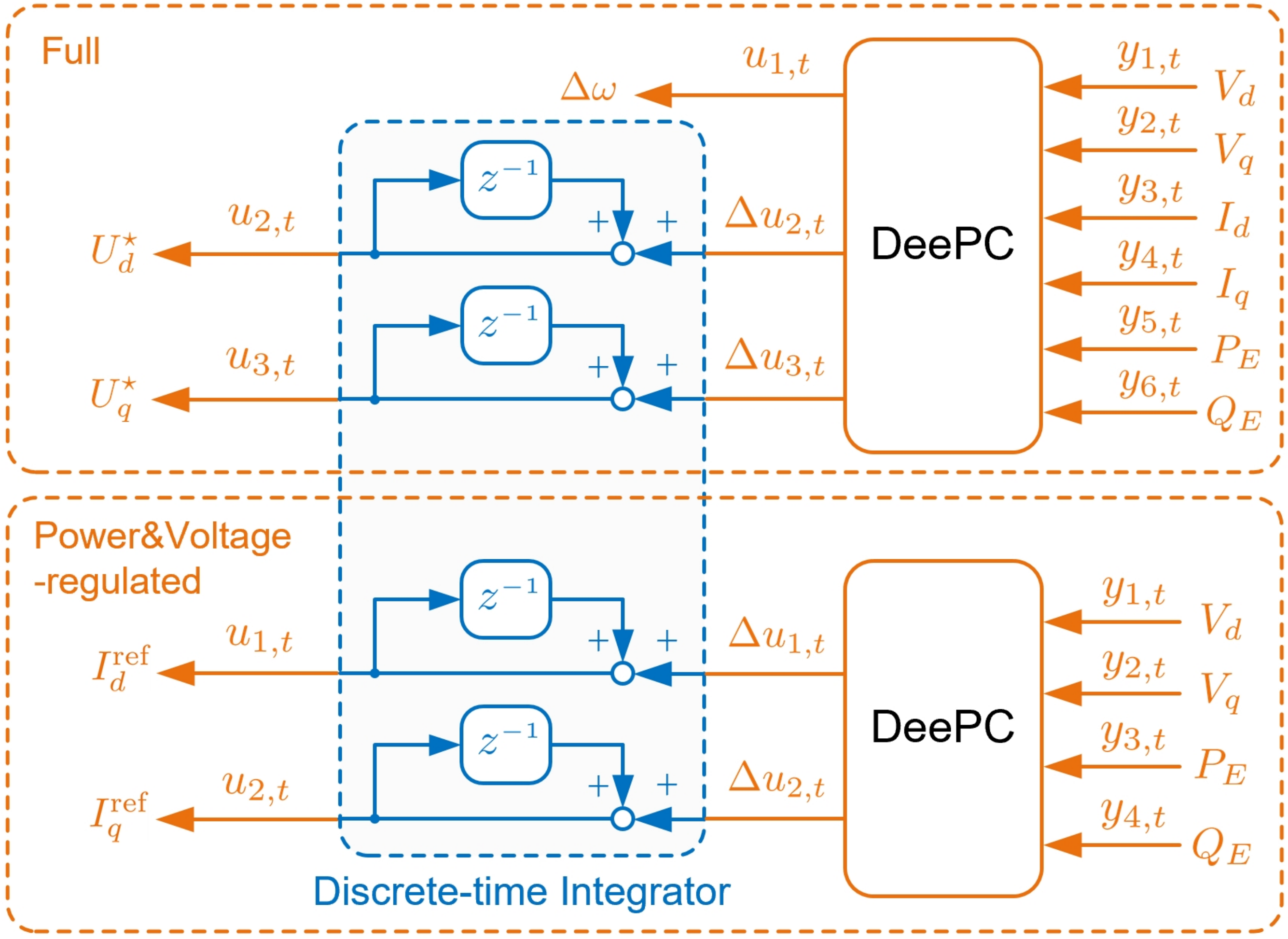}}
	\vspace{-2mm}
        \caption{Control block diagrams of the integral DeePConverters: Full and Power\&Voltage-regulated configurations.}
        \label{integral}
	\vspace{-2mm}
\end{figure}

Fig.~\ref{integral} shows the control block diagrams of the integral form for two configurations: Full and Power\&Voltage-regulated DeePConverters. In both cases, integrators are inserted into the selected control channels. The corresponding formulations for each configuration are detailed below.

\textit{Integral Full DeePConverter:} In this configuration, integrators are inserted in the $u_2$ and $u_3$ channels. The DeePC optimizes the variable set $[u_{1,t},\, \Delta u_{2,t},\, \Delta u_{3,t}]^\top$, where $u_{1,t}$ (frequency) is directly used as the control input without integration, while the other two inputs are computed through accumulation as $u_{2,t} = u_{2,t-1} + \Delta u_{2,t}$ and $u_{3,t} = u_{3,t-1} + \Delta u_{3,t}$.

\textit{Integral Power\&Voltage-regulated DeePConverter:} In this configuration, integrators are inserted in the $u_1$ and $u_2$ channels. The DeePC optimizes the differences $[\Delta u_{1,t},\, \Delta u_{2,t}]^\top$, and then the corresponding control inputs are reconstructed as $u_{1,t} = u_{1,t-1} + \Delta u_{1,t}$ and $u_{2,t} = u_{2,t-1} + \Delta u_{2,t}$.

\begin{figure*}[!t]
 \vspace{0mm}
\centerline{\includegraphics[width=1\linewidth]{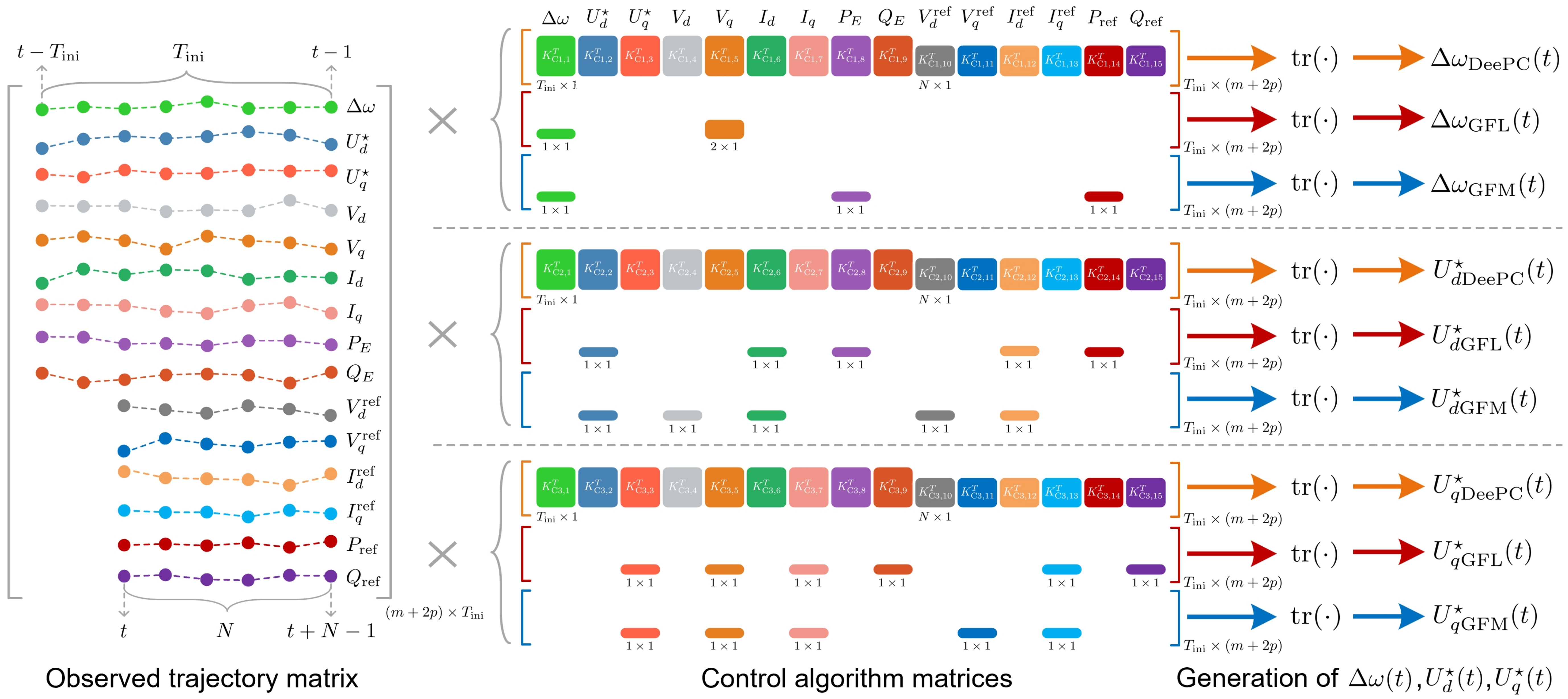}}
	\vspace{-2mm}
        \caption{Control signal generation for three converter types: mapping observed trajectory matrix to $\Delta \omega(t)$, $U_d^\star(t)$, and $U_q^\star(t)$ via control algorithm matrices (with zeros in place of non-existent components), with GFL and GFM matrix elements determined by controller parameters as in~\eqref{eq:PLL1} and~\eqref{eq:swing_dis}, and Full DeePConverter as a generalized structure encompassing both GFL and GFM schemes, with control elements formed from vectors in $K_{\rm C}$ of~\eqref{eq:mapping_u0}.}
        \label{Generalized_structure}
	\vspace{-2mm}
\end{figure*}

\section{Generalized Structure, Behavior Design, and Implementation of Full DeePConverters}
Among the available configurations, the Full DeePConverter serves as a representative example due to its ability to replace all primary control loops. In this section, we illustrate its generalized data-driven control architecture, discuss the behavior specification via cost function design, highlight its robustness to data disturbances, and detail its efficient implementation.

\subsection{Generalized Data-Driven Control Architecture}\label{general}

We now focus on the generalized control structure enabled by the Full DeePConverter and show that it includes both GFL and GFM control schemes within a data-driven formulation. In this configuration, the DeePC controller generates the frequency deviation $\Delta \omega$ and internal voltage references ($U_d^\star$, $U_q^\star$) based on the measured system outputs. These outputs include the $d$- and $q$-axis voltages ($V_d$, $V_q$), $d$- and $q$-axis currents ($I_d$, $I_q$), active and reactive power ($P_E$, $Q_E$), which serve as feedback signals for computing the next control inputs.

Specifically, when the input/output constraints (e.g., current/voltage saturations) are inactive, the optimization problem~\eqref{eq:DeePC} with $h(g) = \|g\|_2^2$ admits a closed-form solution~\cite{huang2021quadratic}, which linearly maps the initial trajectory $[u_{\mathrm{ini}}; y_{\mathrm{ini}}]$ and the reference trajectory $r$ to the first-step future inputs, that is
\begin{equation}\label{eq:mapping_u0}
u_t = \begin{bmatrix} \Delta \omega(t);U_d^{\star}(t);U_q^{\star}(t) \end{bmatrix} = u_0^\star = K_{\rm C} \begin{bmatrix}  u_{\rm ini};y_{\rm ini}; r \end{bmatrix} ,
\end{equation}
where $K_{\rm C} \in \mathbb{R}^{m \times (mT_{\rm ini}+ pT_{\rm ini}+pN)}$ is the control matrix determined by the optimization problem. Notice that $[u_{\rm ini};y_{\rm ini}]$ contains the past length-$T_{\rm ini}$ trajectories of the outputs ($V_d$, $V_q$, $I_d$, $I_q$, $P_E$, and $Q_E$) and the inputs ($\Delta \omega$, $U_d^{\star}$, and $U_q^{\star}$), e.g., $[V_q(t-T_{\rm ini});V_q(t-T_{\rm ini}+1);\dots;V_q(t-1)]$. The vector $r$ contains the reference trajectories corresponding to each output variable, that is, $[V_d^{\mathrm{ref}}, V_q^{\mathrm{ref}}, I_d^{\mathrm{ref}}, I_q^{\mathrm{ref}}, P_{\mathrm{ref}}, Q_{\mathrm{ref}}]$.

\subsubsection{Including GFL Control}
For a GFL converter, its frequency is generated by a PLL whose control law in discrete time can be expressed as
\begin{equation}\label{eq:PLL}
\Delta \omega(t) = K_{\rm P,PLL}V_q(t-1) + \frac{K_{\rm I,PLL}T_{\rm S}}{z-1} V_q(t),
\end{equation}
where $z$ can be considered as the shift operator (i.e., $zx(t) = x(t+1)$), $T_{\rm S}$ is the sampling time, $K_{\rm P,PLL}$ and $K_{\rm I,PLL}$ are the PI parameters of the PLL. We use $V_q(t-1)$ instead of $V_q(t)$ for the proportional term such that the control law is implementable in practice. We further rewrite~\eqref{eq:PLL} as
\begin{IEEEeqnarray}{rCl}\label{eq:PLL1}
\Delta \omega(t) 
&=& \Delta \omega(t-1) + (K_{\rm P,PLL} + K_{\rm I,PLL}T_{\rm S})V_q(t-1) \nonumber\\
&& - K_{\rm P,PLL}V_q(t-2),
\end{IEEEeqnarray}
which is equivalent to the first row of~\eqref{eq:mapping_u0} (with $T_{\rm ini} \ge 2$) by choosing the first row of $K_{\rm C}$ accordingly to map $\Delta \omega(t-1)$, $V_q(t-1)$, and $V_q(t-2)$ (which are elements of $[u_{\rm ini};y_{\rm ini}]$) to $\Delta \omega(t)$. Analogously, one can show that a proper choice of the second (third) row leads to a cascaded power outer loop and current loop control structure, where the second (third) row uses $P_E$, $P_{\rm ref}$, $I_d$, and $I_d^{\mathrm{ref}}$ ($Q_E$, $Q_{\rm ref}$, $I_q$, and $I_q^{\mathrm{ref}}$) to generate the internal voltage reference $U_d^\star$ ($U_q^\star$). Hence, the control structure of Full DeePConverter includes the typical control scheme of a (PLL-based) GFL converter.

\subsubsection{Including GFM Control}
For a GFM converter (e.g., a VSM~\cite{d2015virtual,yang2020placing}), the frequency is generated according to the swing equation whose continuous-time formulation is
\begin{equation}\label{eq:swing_con}
Js\Delta \omega + D \Delta \omega = P_{\rm ref} - P_E ,
\end{equation}
where $J$ is the inertia coefficient and $D$ is the damping coefficient.
The discrete-time control law of~\eqref{eq:swing_con} is
\begin{equation}\label{eq:swing_dis}
\Delta \omega(t) = \frac{P_{\rm ref}T_{\rm S}}{J} + \left( 1 - \frac{DT_{\rm S}}{J} \right) \Delta \omega(t-1) - \frac{T_{\rm S}}{J} P_E(t-1) ,
\end{equation}
which, again, can be captured by the first row of~\eqref{eq:mapping_u0} (with $T_{\rm ini} \ge 1$) by choosing the first row of $K_{\rm C}$ accordingly. Then, one can choose the second (third) row of $K_{\rm C}$ to use $V_d$, $V_d^{\rm{ref}}$, $I_d$, $I_d^{\rm{ref}}$ ($V_q$, $V_q^{\rm{ref}}$, $I_q$, $I_q^{\rm{ref}}$) to generate $U_d^\star$ ($U_q^\star$), acting as the cascaded voltage and current control loops in GFM converters.

In sum, the Full DeePConverter can be viewed as a generalized control structure that includes both GFM and GFL control schemes. As shown in Fig.~\ref{Generalized_structure}, it offers higher structural flexibility by enabling more degrees of freedom in the control matrix, where many entries are fixed to zero in traditional designs. Our framework directly encodes control policies as linear combinations of historical inputs, outputs, and reference trajectories, i.e., $[u_{\rm ini}; y_{\rm ini}; r]$. This formulation enhances the expressiveness of DeePConverter and raises its theoretical performance ceiling. While the structure offers the potential to cover both GFM and GFL schemes, the realization of specific control behaviors ultimately hinges on cost function design. This aspect is detailed in the following section.

\subsection{Behavior Specification via Cost Function Design}

Unlike GFL and GFM converters, whose control structures and synchronization mechanisms are a priori specified based on engineering insights, DeePConverters encode control objectives directly within the cost function of~\eqref{eq:DeePC}. By properly designing the term $\|y - r\|_Q^2$, one can induce desired behaviors.

\subsubsection{General Behavior Specification}
To show how basic control functionalities are embedded, we first rewrite the output cost term in an explicit structured form as
\begin{equation} \label{eq:full_cost_function_qv}
\begin{aligned}
\|y - r\|_Q^2 = &\;
\alpha_1 \left\| 
\begin{bmatrix}
\Delta \omega(t:t+N-1) \\
P_E(t:t+N-1)
\end{bmatrix}
-
\begin{bmatrix}
\boldsymbol{\phi}_N \\
P_{\rm ref} \cdot \mathbf{1}_N
\end{bmatrix}
\right\|_{Q_{P\omega}}^2 \\
&+ \alpha_2 \left\| 
\begin{bmatrix}
V_d(t:t+N-1) \\
Q_E(t:t+N-1)
\end{bmatrix}
-
\begin{bmatrix}
V_d^{\rm ref} \cdot \mathbf{1}_N \\
Q_{\rm ref} \cdot \mathbf{1}_N
\end{bmatrix}
\right\|_{Q_{QV}}^2 \\
&+ \alpha_3 \|V_q(t:t+N-1) - V_q^{\rm ref} \cdot \mathbf{1}_N\|^2
\end{aligned}
\end{equation}
where \( V_d, V_q, Q_E, P_E, \Delta \omega \in \mathbb{R}^N \) denote the predicted trajectories over horizon \( N \);  
\( V_d^{\rm ref}, V_q^{\rm ref}, Q_{\rm ref}, P_{\rm ref} \in \mathbb{R} \) are reference values. The vector \( \boldsymbol{\phi}_N \in \mathbb{R}^N \) is an adjustable offset. The weights \( \alpha_1, \alpha_2, \alpha_3 \) correspond to active-power control, reactive-power/voltage control, and voltage-orientation, respectively;  
\( Q_{P\omega}, Q_{QV} \in \mathbb{R}^{2N \times 2N} \) are coupling matrices.

Here, the converter-side currents \( I_d \) and \( I_q \) are not directly regulated via the cost function; rather, they are included in the output vector \( y \) to facilitate current-related constraints, which will be discussed later. In the remainder of this section, we focus on~\eqref{eq:full_cost_function_qv} and illustrate how different control objectives can be achieved by appropriately tuning the weights (\( \alpha_1 \), \( \alpha_2 \), \( \alpha_3 \)) and designing the coupling matrices \(( Q_{P\omega}, Q_{QV} \)).

\textit{a) Voltage orientation}:  
By setting $V_q^{\rm ref} = 0$ and assigning a sufficiently large weight $\alpha_3$ to penalize the tracking error of $V_q$, the voltage vector aligns with the $d$-axis at steady state.

\textit{b) Voltage magnitude regulation}:  
By setting a large $\alpha_2$ and designing the matrix \( Q_{QV} \), different voltage control modes can be achieved through the following structure. We choose
\begin{equation}\label{qv}
Q_{QV} =
\begin{bmatrix}
\alpha_V & \beta \\
\beta   & \alpha_Q 
\end{bmatrix}\otimes I_N,
\end{equation}
where \( \alpha_V, \alpha_Q \in \{0,1\} \) enable voltage and reactive power tracking, and \( \beta \in \mathbb{R} \) sets the Q-V droop slope. This yields: AC voltage control (\( \alpha_V = 1, \alpha_Q = 0, \beta = 0 \)) for PV behavior; reactive power control (\( \alpha_V = 0, \alpha_Q = 1, \beta = 0 \)) for PQ behavior; and Q-V droop behavior (\( \alpha_V = 1, \alpha_Q = k_{QV}^2, \beta = k_{QV} \)) with droop gain \( k_{QV} \in \mathbb{R}_+ \).

\textit{c) Active power regulation \& synchronization:}  
By setting a large weight $\alpha_1$ and designing the matrix \( Q_{P\omega} \) as
\begin{equation}\label{pw}
Q_{P\omega} =
\begin{bmatrix}
0 & 0 \\
0 & 1
\end{bmatrix} \otimes I_N,
\end{equation}
active power tracking is enforced (i.e., $P_E = P_E^{\rm ref}$ at steady state).  
Once the voltage is properly oriented and its magnitude stabilized, regulating active power naturally leads to synchronization with the grid, since steady-state power exchange and voltage cannot be maintained without phase alignment.

In the above, the converter regulates active power assuming a constant or well-controlled DC voltage. Alternatively, when DC voltage regulation is required~\cite{cvetkovic2014modeling}, one can treat the DC voltage as an output and penalize its tracking error in~\eqref{eq:full_cost_function_qv} in place of active power to achieve DC voltage regulation.

\subsubsection{GFL and GFM Behavior Realization}
As discussed in subsection~\ref{general}, DeePConverters offer a generalized structure including GFL and GFM control. Building on this, we now show how GFL and GFM behaviors can be induced through cost function engineering, by appropriately structuring~\eqref{eq:full_cost_function_qv}.

\textit{a) GFL behavior realization:}  
To approximate the behavior of GFL converters, which typically synchronize with the grid via PLL and track power references with fast dynamics, we design cost terms in~\eqref{eq:full_cost_function_qv} that enforce these objectives.

To encode the synchronization behavior, we assign a sufficiently large weight \(\alpha_3\) to the \(V_q\) tracking term, thereby enforcing \(V_q\) to \(0\). This ensures the rotating reference frame remains aligned with the grid voltage vector, effectively replicating the phase-locking behavior of a PLL in the GFL converters.

The active and reactive power control behaviors are implemented via diagonal coupling matrices defined as
\begin{equation}
Q_{P\omega} =
\begin{bmatrix}
0 & 0 \\
0 & 1
\end{bmatrix} \otimes I_N, \quad
Q_{QV} =
\begin{bmatrix}
0 & 0 \\
0 & 1
\end{bmatrix} \otimes I_N,
\end{equation}
which independently penalize the deviations of \(P_E\) and \(Q_E\) from their respective references. The offset \(\boldsymbol{\phi}_N \in \mathbb{R}^N\) remains unconstrained and is typically set to \(\mathbf{0}_N\) for simplicity, without affecting the active power tracking objectives.

Given that PLL dynamics are usually faster than the active/reactive power loops in the GFL converters, we assign weight priorities to reflect this characteristic: \( \alpha_3 > \alpha_1 = \alpha_2 \).

\begin{table}[!t]
\centering
\caption{Summary of Weight Priorities, Quadratic Forms, and Offset Terms in GFL and GFM Behavior Realization}
\label{tab:GFL_GFM}
\renewcommand{\arraystretch}{1.15}
\setlength{\tabcolsep}{3.5pt}  
\begin{tabular}{@{}lcccc@{}}
\toprule
\textbf{Mode} & \textbf{$\alpha$‐Weights} & {\boldmath$Q_{P\omega}$} & {\boldmath$Q_{QV}$} & {\boldmath$\boldsymbol{\phi}_N$} \\ \midrule
GFL &
$\alpha_3 > \alpha_1 = \alpha_2$ &
$\begin{bmatrix} 0 & 0 \\ 0 & 1 \end{bmatrix}\!\otimes\! I_N$ &
$\begin{bmatrix} 0 & 0 \\ 0 & 1 \end{bmatrix}\!\otimes\! I_N$ &
$\boldsymbol{\phi}_N\!\in\!\mathbb{R}^N$ \\[8pt]
GFM &
$\alpha_2 = \alpha_3 > \alpha_1$ &
$\begin{bmatrix} M^{\!\top}M & M^{\!\top} \\ M & I_N \end{bmatrix}$ &
$\begin{bmatrix} 1 & 0 \\ 0 & 0 \end{bmatrix}\!\otimes\! I_N$ &
$\dfrac{JM^{-1} d_0}{T_{\rm S}}$ \\ \bottomrule
\end{tabular}
\end{table}
\vspace{-0mm}

\textit{b) GFM behavior realization:}  
To emulate the behavior of GFM converters, particularly under VSM control, we construct cost terms in~\eqref{eq:full_cost_function_qv} that enforce both active power-frequency dynamics and voltage regulation.

To reproduce the swing equation in~\eqref{eq:swing_dis}, we design the active power-frequency cost term with
\begin{equation} \label{eq:swing}
Q_{P\omega} =
\begin{bmatrix}
M^{\!\top}M & M^{\!\top} \\
M & I_N
\end{bmatrix}, \quad
\boldsymbol{\phi}_N = \frac{J}{T_{\rm S}} M^{-1} d_0,
\end{equation}
where \( M = D I_N + \frac{J}{T_{\rm S}} \mathcal{D}_N \), with \( D \) and \( J \) denoting the damping coefficient and virtual inertia, respectively, and \( \mathcal{D}_N \in \mathbb{R}^{N \times N} \) being the backward-difference operator. 

With \( d_0 = \begin{bmatrix} \Delta\omega(t-1); 0; \cdots; 0 \end{bmatrix} \), the power–frequency cost term in~\eqref{eq:full_cost_function_qv} reduces to a trajectory-level residual that closely resembles the discrete-time swing equation
\begin{equation*}
    \alpha_1\left\|
    \begin{array}{l}
        \frac{J}{T_{\rm S}} [\Delta \omega(t:t{+}N{-}1)-\Delta \omega(t-1:t{+}N{-}2)]\\
        +D\,\Delta \omega(t:t{+}N{-}1)- P_{\text{ref}} \, \mathbf{1}_N + P_E(t:t{+}N{-}1) 
    \end{array}\hspace{-1mm}
    \right\|^2 \,,
\end{equation*}
which embeds the GFM synchronization dynamics directly into the cost formulation through virtual inertia and damping.

The AC voltage-regulation behavior is captured by the last two terms in~\eqref{eq:full_cost_function_qv}. First, the reactive power and \(d\)-axis voltage tracking is enforced via the block
\begin{equation}
Q_{QV} =
\begin{bmatrix}
1 & 0 \\
0 & 0
\end{bmatrix} \otimes I_N.
\end{equation}
which penalizes deviation from the reference \(V_d^{\rm ref}\) while leaving \(Q_E\) unconstrained. The third term in~\eqref{eq:full_cost_function_qv} further regulates the \(q\)-axis voltage \(V_q\) toward \(V_q^{\rm ref}\), thereby completing the orientation of the AC voltage vector in the \(dq\) frame.

As GFM converters typically feature fast voltage regulation coordinated with comparatively slower active power-frequency control, the weighting parameters are selected to reflect this inherent timescale separation, assigning higher priority to the voltage control objectives: \( \alpha_2 = \alpha_3 > \alpha_1 \). The corresponding settings of weighting coefficients, quadratic forms, and offset terms for the GFL and GFM modes are summarized in Table~\ref{tab:GFL_GFM}.

While GFL and GFM behaviors can be reliably  emulated with improved performance using the above setting (as demonstrated by simulations in the next section), our approach is not limited to these existing modes. By tailoring the cost function to the system needs, DeePConverters enable a systematic and optimal design beyond conventional methods.

\subsection{Inherent Robustness to Data Disturbances}\label{sec:robust}

\subsubsection{Robustness to Measurement Noise and Inaccuracies}
Measurement noise, sensor bias, and parameter drift are commonly observed in power systems, manifesting as additive disturbances in \( u_{\mathrm{ini}}, y_{\mathrm{ini}} \), and as column-wise perturbations in the Hankel matrices \( U_{\rm{P}}\), \(Y_{\rm{P}}\), \(U_{\rm{F}}\), \(Y_{\rm{F}}\). Such imperfections may degrade the accuracy of the behavioral model and consequently the control performance if not properly addressed~\cite{huang2021quadratic}. To enhance robustness against these data corruptions, the DeePC formulation~\eqref{eq:DeePC} introduces three penalty terms:
\begin{itemize}
    \item \textbf{Slack penalties} with \( \lambda_u, \lambda_y \geq 0 \) introduce slack variables to relax the consistency constraints for the recorded past input \( u_{\mathrm{ini}} \) and output \( y_{\mathrm{ini}} \), penalizing deviations from the reconstructed trajectories \( U_{\rm{P}} g \) and \( Y_{\rm{P}} g \), respectively.
    \item \textbf{Regularization penalty} with \( \lambda_g \geq 0 \): typical choices of the regularizer \( h(g) \) include (a) \( h(g) = \|g\|_2^2 \), (b) \( h(g) = \|g\|_1 \), and (c) a projection-based form \( h(g) = \|(I - \Pi)g\|_2^2 \), each offering distinct robustness properties against different uncertainty structures.
\end{itemize}

These penalty terms jointly endow DeePC with robustness guarantees. When \( \lambda_u \) and \( \lambda_y \) exceed noise-dependent thresholds, the consistency constraints act as quadratic barriers that upper-bound the worst-case closed-loop cost under bounded Hankel perturbations~\cite{huang2023robust}. The quadratic regularizer \( \|g\|_2^2 \) corresponds to a distributionally robust formulation over a Frobenius-norm ball, while the \( \ell_1 \) and projection regularizers address column-wise and subspace uncertainties~\cite{markovskydata}. Joint tuning of \( \lambda_u, \lambda_y \), and \( \lambda_g \) yields a conic-quadratic problem that remains tractable under realistic noise and offers a principled trade-off between robustness and performance~\cite{huang2023robust}.

\subsubsection{Robustness to Communication Delays}
Since DeePC constructs its predictor directly from input/output trajectories, fixed or slowly varying communication delays are naturally embedded in the historical data. In discrete time, such delays correspond to deterministic linear phase shifts and therefore do not introduce additional modeling complexity beyond the intrinsic system dynamics. Hence, the data-driven predictor remains valid and the standard DeePC formulation can maintain robust performance without modification. In contrast, rapidly time-varying or stochastic delays break this deterministic structure and remain a challenge for future research.

\subsubsection{Robustness to Nonlinearities}
The regularized DeePC formulation is inherently robust to weak nonlinearities encountered by power converters operating around certain steady-state operating points. In this case, the primary uncertainty arises from mismatches between the linear data-driven representation and the true system dynamics, which can be effectively mitigated by the regularization in the DeePC formulation~\cite{dorfler2022bridging,huang2023robust}. For systems dominated by strong nonlinearities, such as nonlinear loads or switching-induced dynamics, promising extensions include kernel-based DeePC~\cite{huang2023kernelized} and neural-network-based data-driven predictors~\cite{wang2005neural}.

\subsection{Implementation, Constraints, and Computational Aspects}

\subsubsection{Implementation}
The detailed procedure for implementing the Full DeePConverter is summarized in Algorithm~\ref{alg:deeptask}. At each control cycle, the controller collects the past trajectory data, solves the DeePC optimization problem, and applies the resulting control inputs to the system. These steps are repeated at every time step to enable closed-loop operation.

\begin{algorithm}[t]
\caption{\normalsize Implementation of the Full DeePConverter} \label{alg:deeptask}
{\bf Input:} Reference trajectory $r$; data matrices $U_{\rm{P}}$, $Y_{\rm{P}}$, $U_{\rm{F}}$, $Y_{\rm{F}}$; cost matrices $R$, $Q$; regularization weights $\lambda_u$, $\lambda_y$, $\lambda_g$.

\begin{enumerate}[1)]
\item At time $t$, collect recent input/output trajectories of length $T_{\mathrm{ini}}$, e.g., $[\Delta \omega(t{-}T_{\mathrm{ini}}); \ldots; \Delta \omega(t{-}1)]$.
  
\item Construct the stacked vector $[u_{\mathrm{ini}}; y_{\mathrm{ini}}]$ using the most recent input/output measurements.
  
\item Solve the DeePC optimization problem, i.e.,~\eqref{eq:DeePC}, to obtain the optimal control sequence $u^\star \in \mathbb{R}^{mN}$.
  
\item Apply the first-step input $u_0^\star \in \mathbb{R}^m$ to the system, i.e., $u_0^\star = [\Delta \omega(t); U_d^\star(t); U_q^\star(t)]$.
  
\item If control horizon $k > 1$, apply $u_i^\star$ at $t{+}i$ for $i \in [1, k{-}1]$.
  
\item Set $t + k \rightarrow t$ and repeat from Step 1.
\end{enumerate}

{\bf Output:} Real-time closed-loop control inputs $u^\star$.
\end{algorithm}

\subsubsection{Constraint Handling in DeePC}
In~\eqref{eq:DeePC}, input and output constraints can be explicitly incorporated to ensure that the DeePConverters operate within safe and practical limits, analogous to the limiters in conventional converter systems. For example, bounds on the current components \( I_d \) and \( I_q \) help prevent overcurrent, while constraints on active and reactive power (\( P_E \), \( Q_E \)) and frequency deviation (\( \Delta \omega \)) ensure compliance with the grid codes and operational safety. When the DC-link voltage \( V_{\mathrm{dc}} \) is used as a control variable instead of the active power, constraints on \( V_{\mathrm{dc}} \) can also be readily incorporated to keep the voltage within safe limits.

\subsubsection{Efficient Solution of the Optimization Problem}
The optimization problem in~\eqref{eq:DeePC} features a quadratic cost and becomes a constrained quadratic program (QP) when input/output inequality constraints are imposed. Solving such QPs can be computationally intensive on embedded hardware. To address this efficiently, we adopt the Operator Splitting Quadratic Program (OSQP) solver~\cite{stellato10operator}, which is well known for its numerical robustness and real-time performance, making it suited for converter applications with constraint requirements.

Specifically, in scenarios where input/output inequality constraints are inactive (or are intentionally ignored) and the quadratic regularizer $h(g)=\|g\|_2^2$ is used, the DeePC problem simplifies considerably into a regularized least-squares formulation. This setting allows us to compute a batch mapping matrix \(M_g\in\mathbb{R}^{H_c\times(mT_{\mathrm{ini}}+pT_{\mathrm{ini}}+pN)}\), which maps the vector \([u_{\mathrm{ini}};\,y_{\mathrm{ini}};\,r]\) to the corresponding optimal coefficient vector \(g^\star\). The mapping \(M_g\) is obtained by solving the following Karush–Kuhn–Tucker (KKT) condition in batch form
\setlength{\arraycolsep}{3pt}
\begin{equation}\label{eq:kkt_batch}
\raisebox{12pt}{$
\begin{bmatrix}
P & 2\lambda_u U_{\rm{P}}^\top & 2\lambda_y Y_{\rm{P}}^\top \\
U_{\rm{P}} & -I_{mT_{\mathrm{ini}}} & 0 \\
Y_{\rm{P}} & 0 & -I_{pT_{\mathrm{ini}}}
\end{bmatrix}
\begin{bmatrix}
M_g \\
M_{\sigma_u} \\
M_{\sigma_y}
\end{bmatrix}
=
\begin{bmatrix}
0 & 0 & 2Y_{\rm{F}}^\top Q \\
I_{mT_{\mathrm{ini}}} & 0 & 0 \\
0 & I_{pT_{\mathrm{ini}}} & 0
\end{bmatrix},$}
\end{equation}
where \(P = 2\left( \lambda_g I_{H_c} + U_{\rm{F}}^\top R U_{\rm{F}} + Y_{\rm{F}}^\top Q Y_{\rm{F}} \right)
\).

Eq.~\eqref{eq:kkt_batch} directly computes the optimal vector \(g^\star\) by leveraging the initial trajectory and the reference \([u_{\mathrm{ini}}; y_{\mathrm{ini}}; r]\) as feedback signals. The first predicted control input for each control channel is further obtained as
\begin{equation}\label{eq:mapping}
\begin{split}
u_0^\star 
&= U_{\rm{F}}(1:m,:) \, g^\star 
= U_{\rm{F}}(1:m,:) \, M_g \, [u_{\mathrm{ini}}; y_{\mathrm{ini}}; r] \\
&= K_{\mathrm{C}} \, [u_{\mathrm{ini}}; y_{\mathrm{ini}}; r].
\end{split}
\end{equation}
where \( K_{\mathrm{C}} = U_{\rm{F}}(1:m,:) M_g \) denotes the control matrix.

This control law is obtained by solving the KKT condition under inactive constraints. Since the control input is computed via a direct linear mapping of the initial trajectory and the reference, it exhibits significant computational efficiency.

\subsection{Online Adaptation of DeePConverters}\label{adaptive}
The Hankel matrices \( U_{\rm P}, Y_{\rm P}, U_{\rm F}, Y_{\rm F} \) are initially constructed from pre-collected data through persistent excitation inputs, such as band-limited white noise, to sufficiently excite the closed-loop dynamics~\cite{markovskydata}. Under normal operation, the DeePC controller assumes that the system behavior remains consistent with the initial dataset. However, practical scenarios involving network reconfiguration, load variations, or parameter drift may lead to performance degradation. To address these challenges, we introduce two online Hankel matrix adaptation strategies: recursive update and batch reconstruction.

\subsubsection{Recursive Update Strategy}
The recursive update strategy incorporates the latest input/output data into the Hankel matrix at each time step via efficient row/column shift operations, enabling real-time adaptation to system variations. This approach is particularly advantageous for rapidly time-varying systems, where small excitation signals superimposed on optimal control inputs facilitate simultaneous regulation and data acquisition while preserving closed-loop performance.

Let \( U^t = \left[ U_{\rm P}^t; U_{\rm F}^t \right] \in \mathbb{R}^{m(T_{\text{ini}}+N) \times H_c} \) and \( Y^t = \left[ Y_{\rm P}^t; Y_{\rm F}^t \right] \in \mathbb{R}^{p(T_{\text{ini}}+N) \times H_c} \) denote the input and output Hankel matrices at time \(t\). The recursive update mechanism is formulated as:
\begin{equation}\label{eq:recursive}
\begin{aligned}
U^{t+1} &= S_U U^t + \begin{bmatrix}
0_{m(T_{\text{ini}}+N-1) \times H_c} \\
R_U^t S_C + \left[ 0_{m \times (H_c-1)} \quad E_u u_{\text{ini}}^{t+1} \right]
\end{bmatrix} \\
Y^{t+1} &= S_YY^t + \begin{bmatrix}
0_{p(T_{\text{ini}}+N-1) \times H_c} \\
R_Y^t S_C + \left[ 0_{p \times (H_c-1)} \quad E_y y_{\text{ini}}^{t+1} \right]
\end{bmatrix}
\end{aligned}
\end{equation}
where \(S_U\) and \(S_Y\) are upward row-shift operators that shift the Hankel blocks by \(m\) and \(p\) rows with zero padding; \(R_U^t\) and \(R_Y^t\) denote the last \(m\) and \(p\) rows of \(U^t\) and \(Y^t\); \(S_C\) is a one-step left-shift column operator with zero padding; and \(E_u = \left[ 0_{m \times m(T_{\text{ini}}-1)} \ \ I_m \right]\), \(E_y = \left[ 0_{p \times p(T_{\text{ini}}-1)} \ \ I_p \right]\) extract the terminal components of the initial-condition vectors.

The updated submatrices are subsequently extracted as:
\begin{equation}
\begin{aligned}
U_{\rm P}^{t+1} &= P_U U^{t+1}, & \quad U_{\rm F}^{t+1} &= F_U U^{t+1}, \\
Y_{\rm P}^{t+1} &= P_Y Y^{t+1}, & \quad Y_{\rm F}^{t+1} &= F_Y Y^{t+1}
\end{aligned}
\end{equation}
where $P_U = \left[ I_{mT_{\text{ini}}} \quad 0_{mT_{\text{ini}} \times mN} \right]$, $F_U = \left[ 0_{mN \times mT_{\text{ini}}} \quad I_{mN} \right]$, $P_Y = \left[ I_{pT_{\text{ini}}} \quad 0_{pT_{\text{ini}} \times pN} \right]$, and $F_Y = \left[ 0_{pN \times pT_{\text{ini}}} \quad I_{pN} \right]$.

This recursive approach provides real-time adaptability, executing updates at every time step to maintain accurate system representation under rapidly varying conditions. The trade-off, however, involves the requirement for persistent low-amplitude excitation to preserve rank conditions of the Hankel matrices.

\subsubsection{Batch Reconstruction Strategy}
In contrast to the recursive approach, the batch reconstruction strategy periodically reconstructs the entire Hankel matrix at designated update instants $t_k$. These updates can be triggered by fixed time intervals or performance degradation, such as excessive tracking errors or constraint violations. This method accumulates input/output data over an adjustable window and performs comprehensive matrix reconstruction in a single operation. The batch update at time $t_k$ is implemented as:
\begin{equation}\label{eq:batch}
\left[ U_{\rm P}^{t_k}; Y_{\rm P}^{t_k}; U_{\rm F}^{t_k}; Y_{\rm F}^{t_k} \right] = \mathscr{H}_{T_{\text{ini}}+N}(\{u_\tau,y_\tau\}_{\tau=t_k-T_{\text{batch}}}^{t_k-1})
\end{equation}
where $T_{\text{batch}}$ denotes the adjustable data window length. Setting it shorter than the nominal data length $T$ during transients accelerates the model adaptation process.

This strategy is characterized by two principal advantages: excitation is only required during the finite data-collection intervals, and Hankel matrix updates occur at a significantly reduced frequency compared to the recursive approach. These attributes render it well-suited for systems with slowly-evolving dynamics or for operational scenarios where persistent excitation or frequent matrix updates are undesirable.

The choice between recursive update and batch reconstruction involves a trade-off between adaptation speed and excitation level. The former provides instantaneous adaptation at the cost of persistent excitation and frequent matrix updates, whereas the latter relaxes excitation requirements at the expense of response latency. The selection should take into account the system’s variability, excitation feasibility, and performance requirements. Both strategies endow DeePC controllers with online adaptability to time-varying dynamics.

\section{Simulation and Experimental Validation}
In this section, we validate the effectiveness and control performance of the proposed DeePConverter framework using high-fidelity simulations and experimental validation. Output measurement noise is incorporated, whereas control input noise is not explicitly modeled. Therefore, the input slack variable $\sigma_{u}$ is omitted. Throughout the validation, all the DeePConverters employ the regularization term $h(g)=\|g\|_2^2$.

\subsection{Simulation of DeePConverter under GFL and GFM Modes}\label{emulate}

We first evaluate the performance of the DeePConverters through detailed simulations on the single-converter system shown in Fig.~\ref{Fig_converter_1}. The system parameters, along with the DeePC-related configurations, are provided in Appendix~\ref{Appendix A}. Two implementations are tested: DeePConverter~1 and DeePConverter~2, corresponding to GFL and GFM operating modes, respectively. Their behaviors are realized by setting the coupling matrices \( Q_{P\omega} \) and \( Q_{PV} \) according to Table~\ref{tab:GFL_GFM}. All DeePConverters adopt the Integral Full version illustrated in Fig.~\ref{integral}, with Hankel matrix data collected prior to \(t=0\). We also test conventional GFL and GFM converters, as well as the GFL and GFM emulations based on Integral SPC (the integral version of subspace predictive control (SPC)~\cite{favoreel1999spc}, which constitutes an indirect data-driven control approach) for comparison.

\subsubsection{Time-Domain Responses}

\begin{figure}[!t]
 \vspace{0mm}
\centerline{\includegraphics[width=1\linewidth]{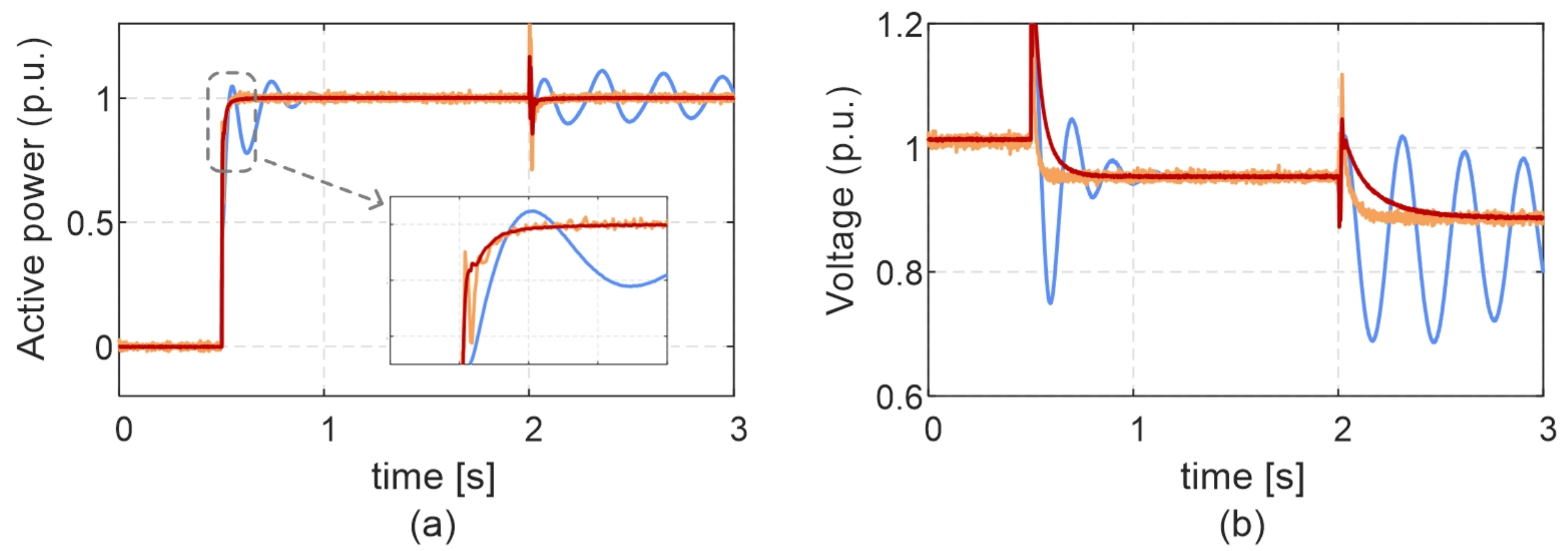}}
	\vspace{-2mm}
        \caption{Time-domain responses of the single-converter system with an initial \(\mathrm{SCR} = 2\). At \(t = 0.5\,\mathrm{s}\), the active power reference steps from \(0\) to \(1\,\mathrm{p.u.}\); at \(t = 2.0\,\mathrm{s}\), the \(\mathrm{SCR}\) drops to \(1.67\).  
        \textcolor[HTML]{C00000}{\textbf{---}}: DeePConverter~1;  
        \textcolor[HTML]{5B8FF7}{\textbf{---}}: GFL converter;  
        \textcolor[HTML]{F7A25B}{\textbf{---}}: GFL emulation based on Integral SPC.}
        \label{GFL}
	    \vspace{-2mm}
\end{figure}

\begin{figure}[!t]
 \vspace{-0mm}
\centerline{\includegraphics[width=1\linewidth]{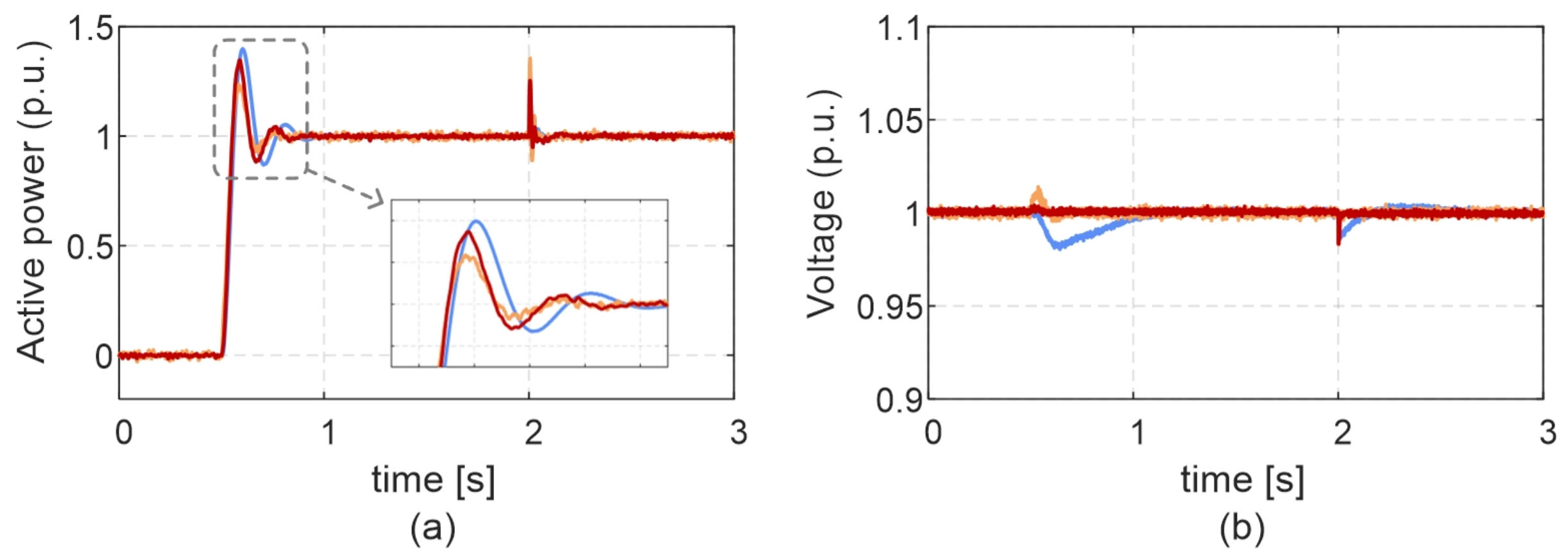}}
	\vspace{-2mm}
        \caption{Time-domain responses of the single-converter system with \(\mathrm{SCR} = 10\). At \(t = 0.5\,\mathrm{s}\), the active power reference steps from \(0\) to \(1\,\mathrm{p.u.}\); at \(t = 2.0\,\mathrm{s}\), the grid voltage drops by \(0.05\,\mathrm{p.u.}\).  
        \textcolor[HTML]{C00000}{\textbf{---}}: DeePConverter~2;  
        \textcolor[HTML]{5B8FF7}{\textbf{---}}: GFM converter;  
        \textcolor[HTML]{F7A25B}{\textbf{---}}: GFM emulation based on Integral SPC.}
        \label{GFM}
	    \vspace{-2mm}
\end{figure}

\begin{figure}[!t]
\centerline{\includegraphics[width=1\linewidth]{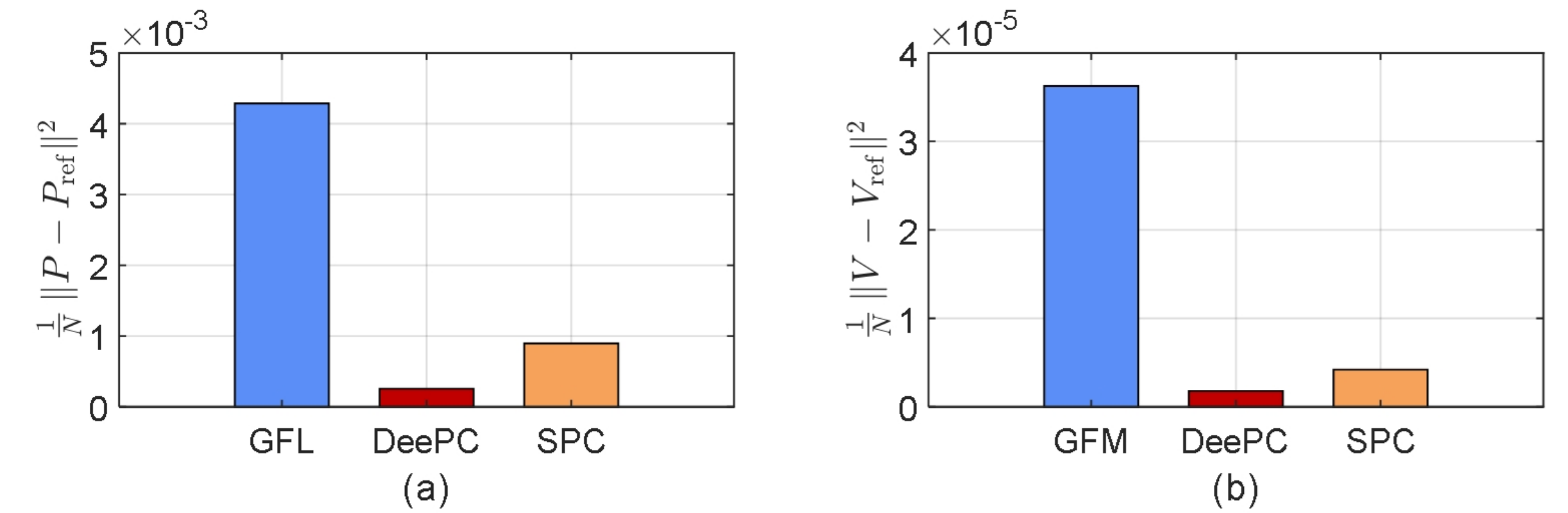}}
    	\vspace{-2mm}
        \caption{Averaged tracking error for the three control strategies: (a) active power tracking in GFL mode; (b) voltage regulation in GFM mode. The metric is computed over the \(0.5\)--\(2.5\,\mathrm{s}\) interval shown in Figs.~\ref{GFL} and~\ref{GFM}, respectively, and averaged over \(20\) trials with independent random measurement noise.}
        \label{fig:quantitative_comparison}
        \vspace{-2mm}
\end{figure}

\begin{figure*}[!t]
\vspace{-5mm} 
\centerline{\includegraphics[width=0.95\linewidth]{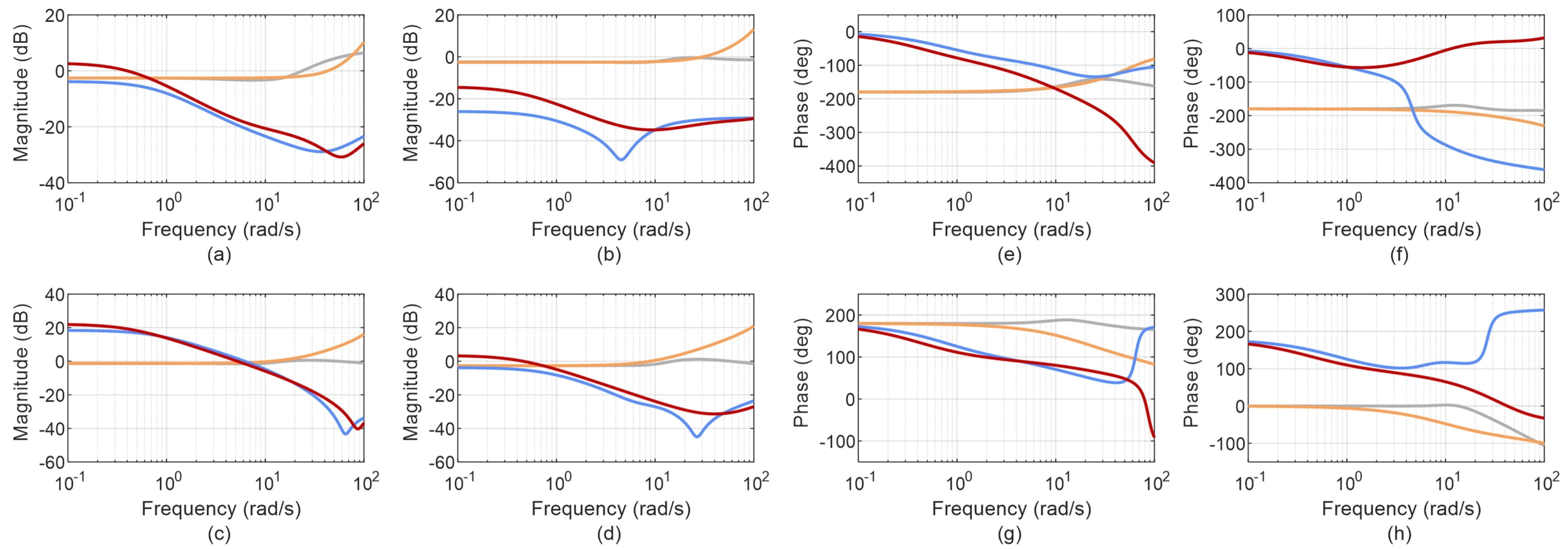}}
    	\vspace{-3mm}    
        \caption{Bode plots of the impedance/admittance matrix elements for four control strategies.
        Subplots (a)--(d) show the magnitude characteristics, and (e)--(h) show the corresponding phase characteristics,
        for the $(1,1)$ to $(2,2)$ matrix elements respectively.
        For GFL mode, admittance-based models are considered:
        \textcolor[HTML]{B0B0B0}{\textbf{---}}: $\mathbf{Y}_{\mathrm{GFL}}(s)$ and
        \textcolor[HTML]{F7A25B}{\textbf{---}}: $\mathbf{Y}_{\mathrm{DeePConverter1}}(s) $.
        For GFM mode, impedance-based models are considered:
        \textcolor[HTML]{5B8FF7}{\textbf{---}}: $\mathbf{Z}_{\mathrm{GFM}}(s)$ and
        \textcolor[HTML]{C00000}{\textbf{---}}: $\mathbf{Z}_{\mathrm{DeePConverter2}}(s)$.}
        \vspace{-2mm}  
        \label{fig:bode_plot}
\end{figure*}

\begin{figure}[!t]
\centerline{\includegraphics[width=0.7\linewidth]{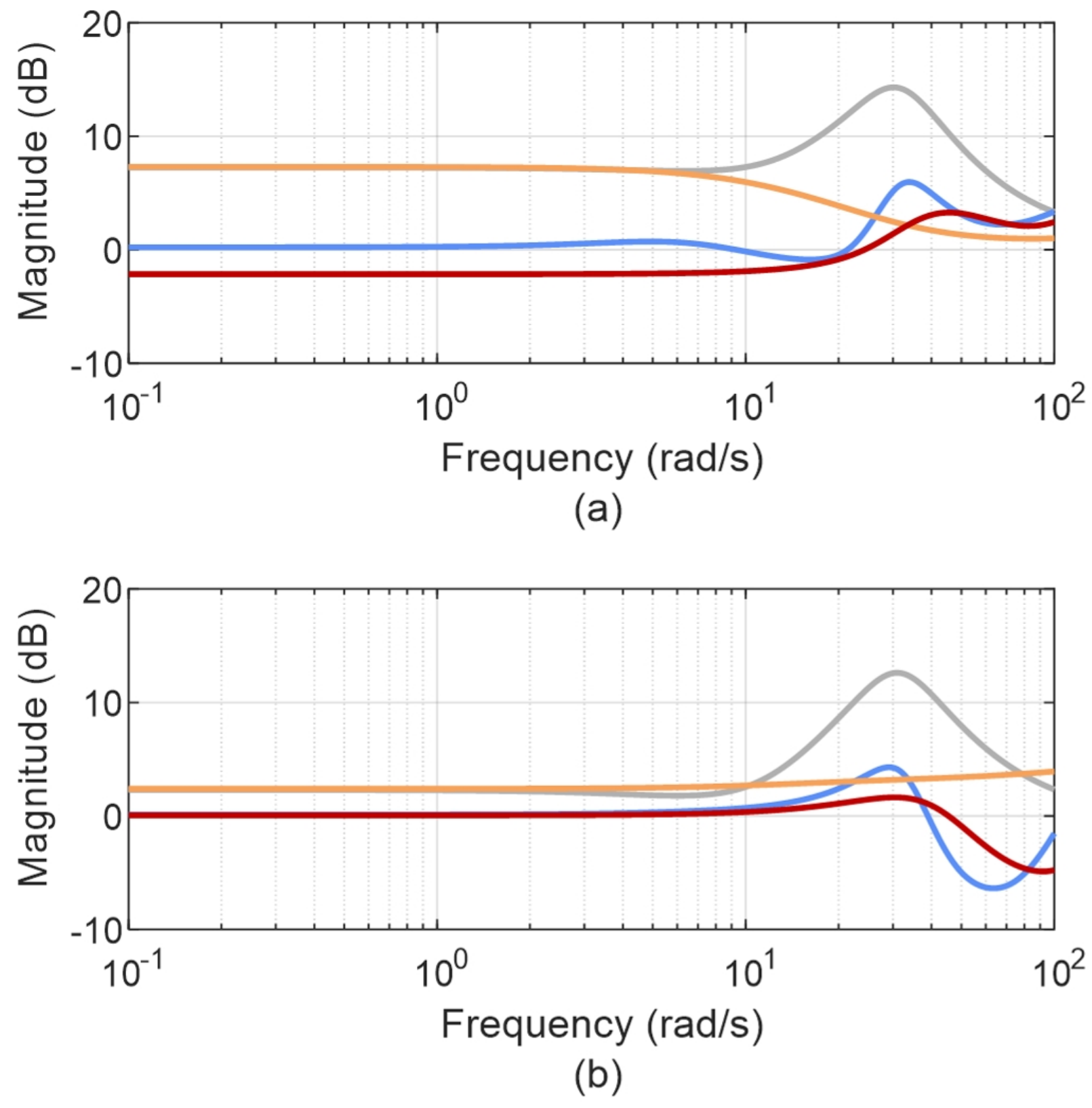}}
        \vspace{-2mm}
        \caption{Maximum singular value curves of the sensitivity and complementary sensitivity functions for four control strategies: (a) $\sigma_{\max}(S(j\omega))$ and (b) $\sigma_{\max}(T(j\omega))$.
        \textcolor[HTML]{B0B0B0}{\textbf{---}}: GFL converter;
        \textcolor[HTML]{5B8FF7}{\textbf{---}}: GFM converter;
        \textcolor[HTML]{F7A25B}{\textbf{---}}: DeePConverter~1;
        \textcolor[HTML]{C00000}{\textbf{---}}: DeePConverter~2.}
        \vspace{-2mm}
    \label{fig:sensitivity}
\end{figure}

Fig.~\ref{GFL} shows the time-domain responses of DeePConverter~1 versus the conventional GFL converter and the SPC-based GFL emulation. DeePConverter~1 successfully emulates the fast active power tracking behavior of a GFL converter. Notably, when the short-circuit ratio (SCR) drops, its power response remains smooth and stable, demonstrating superior performance. In contrast, the conventional GFL converter exhibits significant oscillations following both active power steps and SCR variations, while the SPC-based GFL emulation shows slight sustained oscillations during power tracking and larger active power spikes following SCR changes. Fig.~\ref{GFM} compares the time-domain responses of DeePConverter~2 with the conventional GFM converter and the SPC-based GFM emulation. Both DeePConverter~2 and the SPC-based emulation exhibit inertial and damping characteristics analogous to the swing dynamics of SGs, with DeePConverter~2 providing more pronounced transient power support. Furthermore, during active power steps and grid voltage sags, DeePConverter~2 achieves a faster recovery of voltage magnitude compared to both benchmarks, underscoring its enhanced voltage source behavior. Fig.~\ref{fig:quantitative_comparison} further quantifies the performance of these control strategies by presenting the averaged tracking error of active power in GFL mode and of voltage magnitude in GFM mode. The metric is averaged over \(20\) trials with independent random measurement noise, demonstrating that DeePConverters outperform both the
conventional converters and the SPC-based emulations in terms of active power tracking and voltage regulation.

\subsubsection{Frequency-Domain Characteristics}
Around the steady-state equilibrium point where inequality constraints are inactive, the DeePConverter admits a closed-form solution, which allows the system dynamics to be locally linearized and represented by an equivalent discrete-time linear model. Based on this locally linearized representation, the frequency-domain characteristics are evaluated, as shown in Fig.~\ref{fig:bode_plot}. The Bode plots reveal that DeePConverter~1, operating in GFL mode, well matches the frequency-domain behavior of the conventional GFL converter in both magnitude and phase. 
Similarly, DeePConverter~2 operating in GFM mode emulates the frequency-domain behavior of the conventional GFM converter, particularly in the low-frequency range, and exhibits a better phase response with reduced phase lag. These results show that DeePConverters successfully capture the frequency-domain characteristics of the GFL and GFM strategies.

Additionally, Fig.~\ref{fig:sensitivity} presents the maximum singular value curves, $\sigma_{\max}(S(j\omega))$ and $\sigma_{\max}(T(j\omega))$, for the four control strategies. The $\sigma_{\max}(S(j\omega))$ curves for DeePConverter~1 and 2 exhibit significantly lower peak values compared to those of the conventional GFL and GFM converters, indicating enhanced disturbance rejection. Meanwhile, the $\sigma_{\max}(T(j\omega))$ curves exhibit smoother responses with reduced peaks, indicating improved tracking performance and enhanced robust stability. These frequency-domain observations are consistent with the time-domain simulation results, confirming that the DeePConverters not only emulate the desired GFL and GFM behaviors but also offer superior performance in terms of disturbance rejection, tracking performance, and robust stability.

\subsubsection{Real-Time Feasibility}
The sampling and control period in our simulation is set to \(1 \, \mathrm{ms}\). The Hankel matrix is constructed from \(T = 700\) data points and occupies only approximately \(432 \, \mathrm{kB}\) of memory. On an AMD Ryzen~7~7800X3D processor with \(48 \, \mathrm{GB}\) of random-access memory, solving the constrained QP problem in~\eqref{eq:DeePC_integral} with a quadratic regularizer via the OSQP solver requires about \(5.5 \, \mathrm{ms}\), maintaining a manageable computational burden compared to the SPC method (which requires about \(2.1 \, \mathrm{ms}\)) while achieving superior control performance. Hence, the DeePC method can be implemented in real time by choosing a control horizon \(k > 5\). When the input/output constraints are inactive, the closed-form solution in~\eqref{eq:mapping} can be adopted to achieve faster calculations, which only needs about \(0.001\, \mathrm{ms}\) to obtain the optimal control input.

In summary, the simulation results demonstrate that DeePConverters successfully emulate the GFL and GFM behaviors with enhanced performance. They outperform both conventional and SPC-based methods in power tracking, disturbance rejection, and voltage support, while maintaining real-time feasibility with moderate computational requirements.

\subsection{Transient Performance Tests of the DeePConverter}\label{transient}
We then evaluate the transient performance of the DeePConverters under large disturbances through detailed simulations. Here we focus on testing the performance of active power and voltage regulations. To make a fair comparison with the traditional GFL and GFM control under current limitations, we employ the Integral Self-synchronized version of DeePConverters as shown in Fig.~\ref{Fig_converter_1}, where the current control loops are reserved for current limiting and the upper and lower bounds of \( I_{dq}^{\mathrm{ref}} \) are set to \(\pm 1.2\), the same as those in GFL and GFM control. The system parameters and DeePC-related configurations are provided in Appendix~\ref{Appendix A}.

\subsubsection{Voltage Sag Tests}
\begin{figure}[!t]
\vspace{-4mm}
\centerline{\includegraphics[width=1\linewidth]{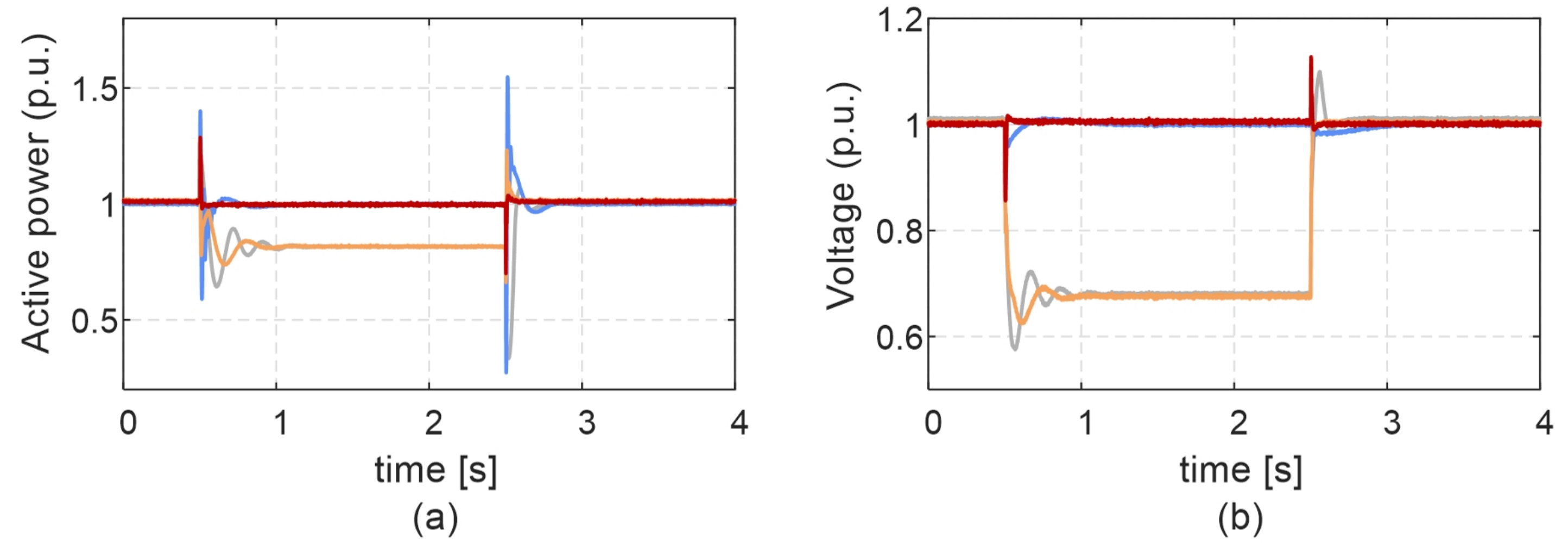}}
        \vspace{-2mm}
        \caption{Time-domain responses of the single-converter system. At \( t = 0.5\,\mathrm{s} \), the grid voltage drops to \(0.7\,\mathrm{p.u.}\) and recovers at \( t = 2.5\,\mathrm{s} \). 
        \textcolor[HTML]{B0B0B0}{\textbf{---}}: GFL converter;
        \textcolor[HTML]{5B8FF7}{\textbf{---}}: GFM converter;
        \textcolor[HTML]{F7A25B}{\textbf{---}}: DeePConverter~PQ;
        \textcolor[HTML]{C00000}{\textbf{---}}: DeePConverter~PV.}
        \vspace{-2mm} 
    \label{fig:voltagedip0.7}
\end{figure}

\begin{figure}[!t]
\centerline{\includegraphics[width=1\linewidth]{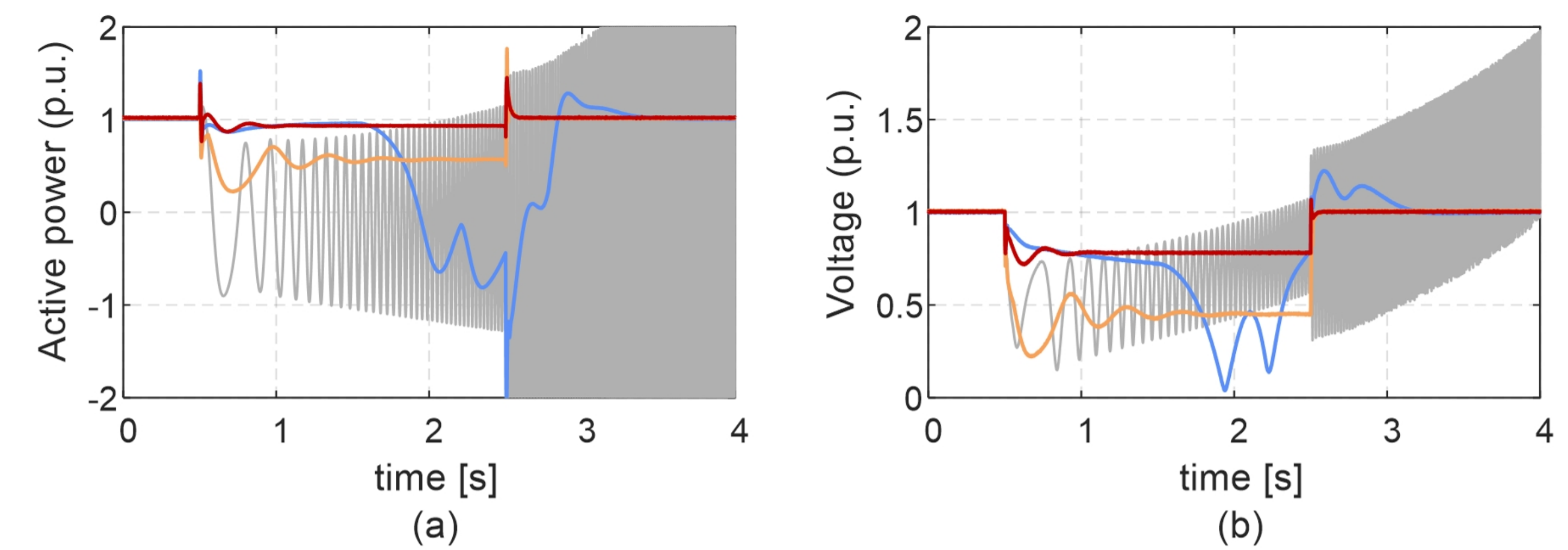}}
        \vspace{-2mm}   
        \caption{Time-domain responses of the single-converter system. At \( t = 0.5\,\mathrm{s} \), the grid voltage drops to \(0.5\,\mathrm{p.u.}\) and recovers at \( t = 2.5\,\mathrm{s} \). 
        \textcolor[HTML]{B0B0B0}{\textbf{---}}: GFL converter;
        \textcolor[HTML]{5B8FF7}{\textbf{---}}: GFM converter;
        \textcolor[HTML]{F7A25B}{\textbf{---}}: DeePConverter~PQ;
        \textcolor[HTML]{C00000}{\textbf{---}}: DeePConverter~PV.}
        \vspace{-2mm} 
    \label{fig:voltagedip0.5}
\end{figure}

The transient performance is first evaluated under severe voltage sag conditions, with conventional GFL and GFM converters included for comparison. The control parameters of the GFL and GFM converters follow the settings in the previous subsection. First, a voltage sag to \(0.7\,\mathrm{p.u.}\) is applied at \(t = 0.5\,\mathrm{s}\) and cleared at \(t = 2.5\,\mathrm{s}\). As shown in Fig.~\ref{fig:voltagedip0.7}, both DeePConverter variants (PQ and PV) maintain stable operation under the fault and converge to steady states faster than their conventional counterparts. A deeper voltage sag to \(0.5\,\mathrm{p.u.}\) is then applied to further test the performance. Fig.~\ref{fig:voltagedip0.5} shows that the GFL converter fails to maintain stable operation and does not recover following the fault clearance, which can be attributed to the transient loss of synchronization under severe voltage sags~\cite{he2021pll}. The GFM converter also experiences a temporary loss of synchronism during the voltage sag, resulting in large deviations in active power and terminal voltage, and then recovers to stable operation after the fault clearance. This transient loss of synchronism is caused by the activation of current limiting under severe voltage sags and has been widely reported for GFM converters in many research works such as~\cite{huang2017transient,shen2020transient}. In contrast, both DeePConverter variants remain stable throughout the entire disturbance and recovery process, showing enhanced transient performance under severe voltage sags.

\subsubsection{Frequency Disturbance Tests}
The performance under frequency disturbances is then evaluated. To assess the inertia and damping characteristics, three DeePConverter configurations are considered: a basic PQ configuration identical to that used in the voltage sag tests, and two enhanced configurations, namely DeePConverter~PQ1 (\(J=1\), \(D=12.5\)) and DeePConverter~PQ2 (\(J=1\), \(D=25\)) (with all other settings identical to the basic PQ configuration), which emulate swing-equation dynamics by configuring \( Q_{P\omega} \) according to~\eqref{eq:swing}, thereby providing inertial and primary frequency response.

\begin{figure}[!t]
\centerline{\includegraphics[width=1\linewidth]{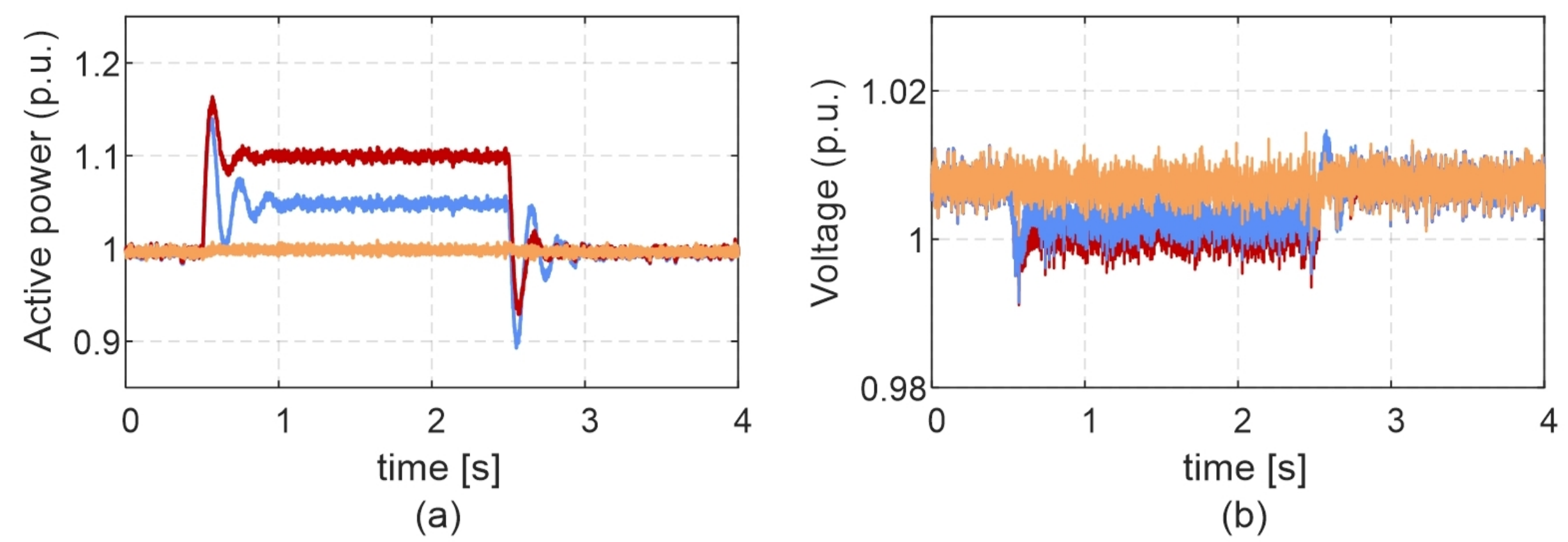}}
        \vspace{-2mm}   
        \caption{Time-domain responses of the single-converter system to a $0.2\,\mathrm{Hz}$ decrease in the grid frequency applied at $t = 0.5\,\mathrm{s}$ and removed at $t = 2.5\,\mathrm{s}$.  
        \textcolor[HTML]{F7A25B}{\textbf{---}}: DeePConverter~PQ (without inertia and damping);
        \textcolor[HTML]{5B8FF7}{\textbf{---}}: DeePConverter~PQ1 (\(J=1\), \(D=12.5\));
        \textcolor[HTML]{C00000}{\textbf{---}}: DeePConverter~PQ2 (\(J=1\), \(D=25\)).}
        \vspace{-2mm} 
    \label{fig:frequency variation}
\end{figure}

A grid frequency step decrease from \(50\,\mathrm{Hz}\) to \(49.8\,\mathrm{Hz}\) is applied at \(t = 0.5\,\mathrm{s}\) and recovers at \(t = 2.0\,\mathrm{s}\). The corresponding time-domain responses are shown in Fig.~\ref{fig:frequency variation}. All three configurations remain stable throughout the entire disturbance. The basic PQ configuration exhibits no active power response to the frequency deviation, whereas both PQ1 and PQ2 provide inertial support and primary frequency regulation. Consistent with the configured damping parameters, the active power of PQ2 stabilizes at approximately \(1.1\,\mathrm{p.u.}\) during the frequency nadir, while that of PQ1 stabilizes at approximately \(1.05\,\mathrm{p.u.}\). These results indicate that the DeePConverter can maintain stable operation under large frequency disturbances while allowing flexible tuning of inertia and damping characteristics.

\subsection{Adaptive Capability Tests of the DeePConverter}\label{adaptive_test}
We evaluate the online adaptive capability of the DeePConverter under changes in external grid operating conditions and variations in internal converter control parameters. In these tests, the DeePConverters adopt the Integral Power-regulated version shown in Fig.~\ref{Fig_converter_1}, with system parameters and DeePC-related configurations provided in Appendix~\ref{Appendix A}. When system changes render the pre-collected data unrepresentative, the recursive update strategy or the batch reconstruction strategy described in Section~\ref{adaptive} can be employed to update the Hankel matrices online and maintain control performance.

\subsubsection{Operating Condition Change}
\begin{figure}[!t]
    \centering
    \vspace{-4mm}
    \includegraphics[width=0.85\linewidth]{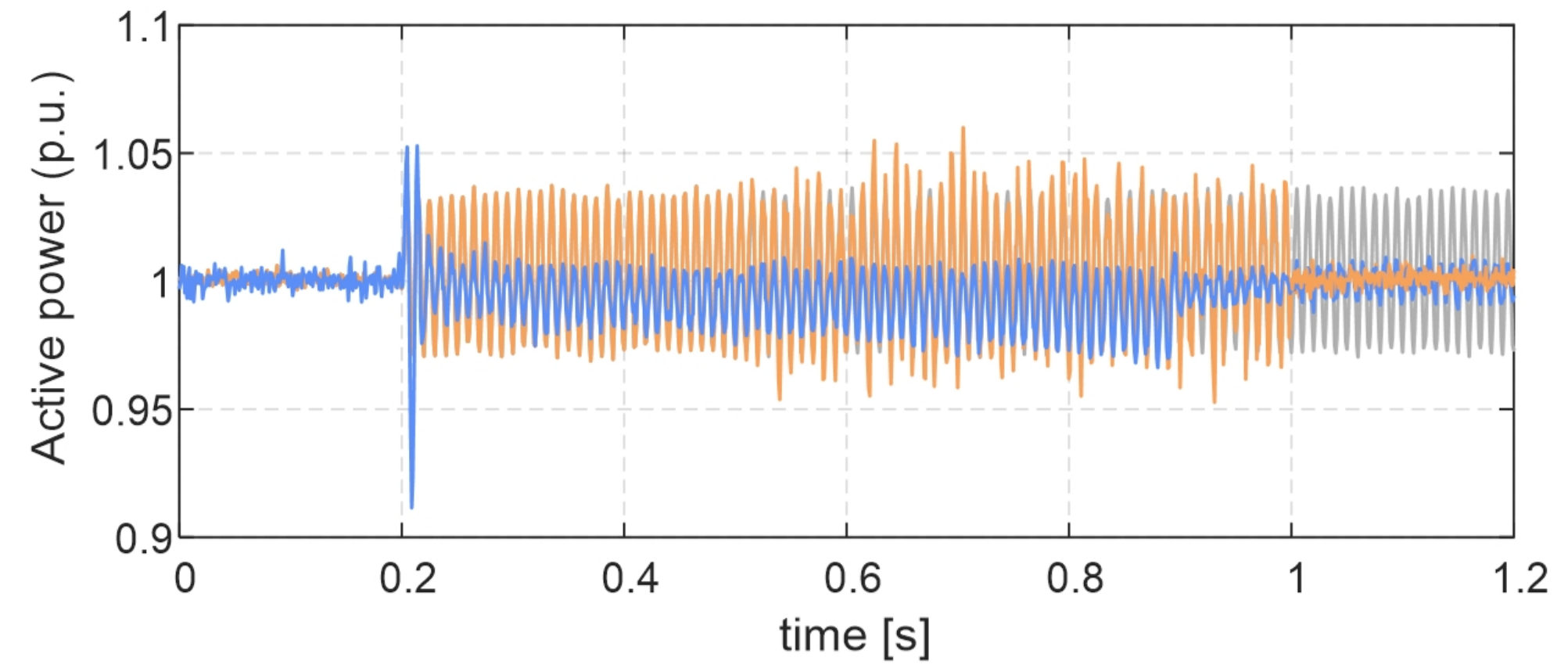}
        \vspace{-2mm}
        \caption{Time-domain responses under a voltage imbalance at \(t = 0.2\,\mathrm{s}\), comparing non-adaptive, recursive update, and batch reconstruction strategies.
        \textcolor[HTML]{B0B0B0}{\textbf{---}}: non-adaptive;
        \textcolor[HTML]{5B8FF7}{\textbf{---}}: recursive update;
        \textcolor[HTML]{F7A25B}{\textbf{---}}: batch reconstruction.}
        \vspace{-2mm}
    \label{fig:unbalanced}
\end{figure}

We first test the adaptability to a change in grid operating conditions by considering an unbalanced voltage deviation. A phase-A voltage change to \(0.95\,\mathrm{p.u.}\) is applied to the single-converter system at \(t = 0.2\,\mathrm{s}\). The time-domain responses under three strategies are shown in Fig.~\ref{fig:unbalanced}. The non-adaptive DeePConverter exhibits slight power oscillations after the fault, due to the voltage imbalance and the fact that the original Hankel matrix no longer accurately represents the altered system dynamics. In contrast, the DeePConverter with the recursive update strategy, which persistently injects low-amplitude excitation signals and updates the Hankel matrix at each time step according to~\eqref{eq:recursive}, effectively adapts to the voltage imbalance and mitigates the power oscillations. The DeePConverter employing the batch reconstruction strategy adapts to the operating condition change by injecting additional excitation signals from \(t = 0.5\,\mathrm{s}\) to \(t = 1.0\,\mathrm{s}\) to collect the fresh input/output data, after which an updated Hankel matrix is reconstructed at \(t = 1.0\,\mathrm{s}\) using a shortened data window (\(T_{\text{batch}} = 500\) instead of the nominal data length \(T = 700\)) for faster adaptation, following~\eqref{eq:batch}. The active power is then well regulated after this update. These results validate the effectiveness of both adaptive strategies in handling changes in operating conditions.

\begin{figure}[!t]
    \centering
    \includegraphics[width=0.85\linewidth]{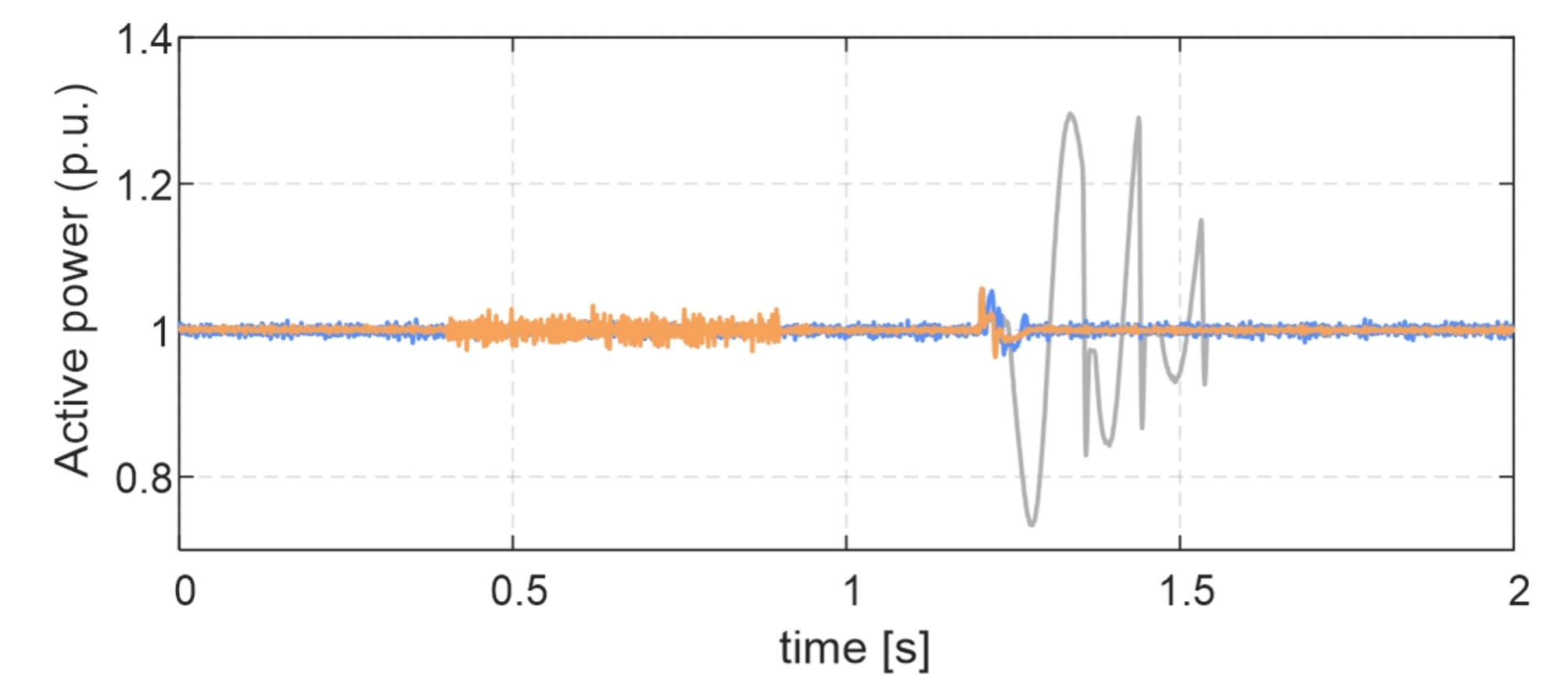}
        \vspace{-2mm}
        \caption{Time-domain responses under a PLL bandwidth change at \(t = 0.2\,\mathrm{s}\) followed by a \(0.1\,\mathrm{p.u.}\) grid voltage sag occurring at \(t = 1.2\,\mathrm{s}\) with a duration of \(0.02\,\mathrm{s}\), comparing non-adaptive, recursive update, and batch reconstruction strategies.
        \textcolor[HTML]{B0B0B0}{\textbf{---}}: non-adaptive;
        \textcolor[HTML]{5B8FF7}{\textbf{---}}: recursive update;
        \textcolor[HTML]{F7A25B}{\textbf{---}}: batch reconstruction.}
        \vspace{-2mm}
    \label{fig:parameter}
\end{figure}

\subsubsection{Parameter Variation}
We further examine the adaptability of the DeePConverter to variations in converter control parameters, aiming to assess its capability to handle, for instance, control switching that is commonly encountered in power converters. For simplicity, the DeePC controller replaces the power control loop, such that the PLL, inner control loops, and the power grid collectively constitute the unknown system dynamics to be learned from data. Consequently, changes in these control parameters may alter the system behavior perceived by the DeePC controller and lead to a mismatch with the pre-collected Hankel matrices.

We consider a step change in the PLL bandwidth from \(10\,\mathrm{rad/s}\) to \(60\,\mathrm{rad/s}\) at \(t = 0.2\,\mathrm{s}\), which emulates, for instance, the impact of an adaptive PLL. Then, a grid voltage sag of \(0.1\,\mathrm{p.u.}\) with a duration of \(0.02\,\mathrm{s}\) occurs at \(t = 1.2\,\mathrm{s}\), and the corresponding responses are shown in Fig.~\ref{fig:parameter}. The non-adaptive DeePConverter exhibits degraded performance with oscillations following the voltage disturbance, as the original Hankel matrix no longer represents the updated system dynamics after the parameter change. In contrast, the recursive update strategy continuously updates the Hankel matrix using newly measured data, enabling fast convergence. The batch reconstruction strategy also restores satisfactory performance by injecting excitation noise from \(t = 0.4\,\mathrm{s}\) to \(t = 0.9\,\mathrm{s}\) and reconstructing the Hankel matrix at \(t = 0.9\,\mathrm{s}\) according to~\eqref{eq:batch}. These results indicate that both adaptive strategies can effectively maintain satisfactory control performance under significant parameter variations.

\subsection{Multi-Converter System Tests of the DeePConverter}

 \begin{figure}[!t]
 \vspace{-4mm}
\centerline{\includegraphics[width=0.8\linewidth]{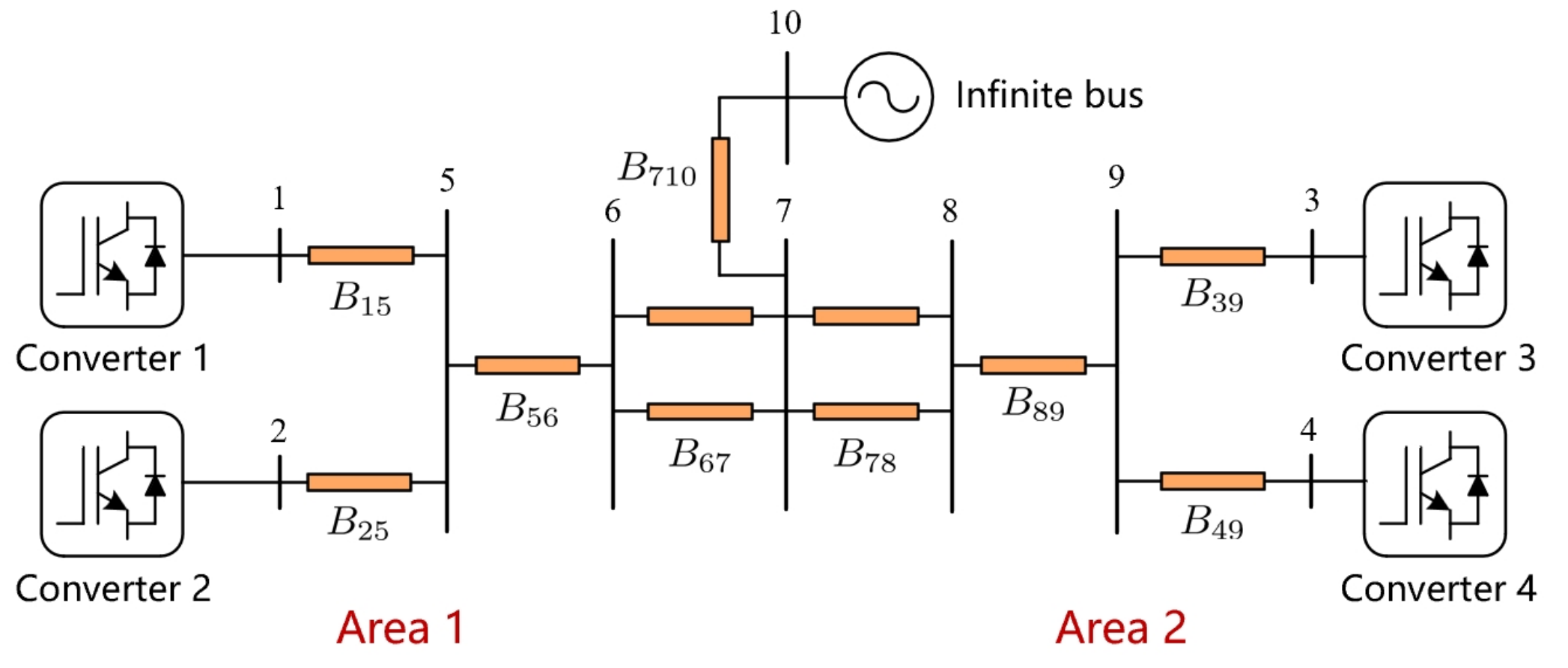}}
	\vspace{-2mm}
        \caption{A two-area test system.}
        \label{two area}
	\vspace{-2mm}
\end{figure}

 \begin{figure}[!t]
 \vspace{0mm}
\centerline{\includegraphics[width=0.85\linewidth]{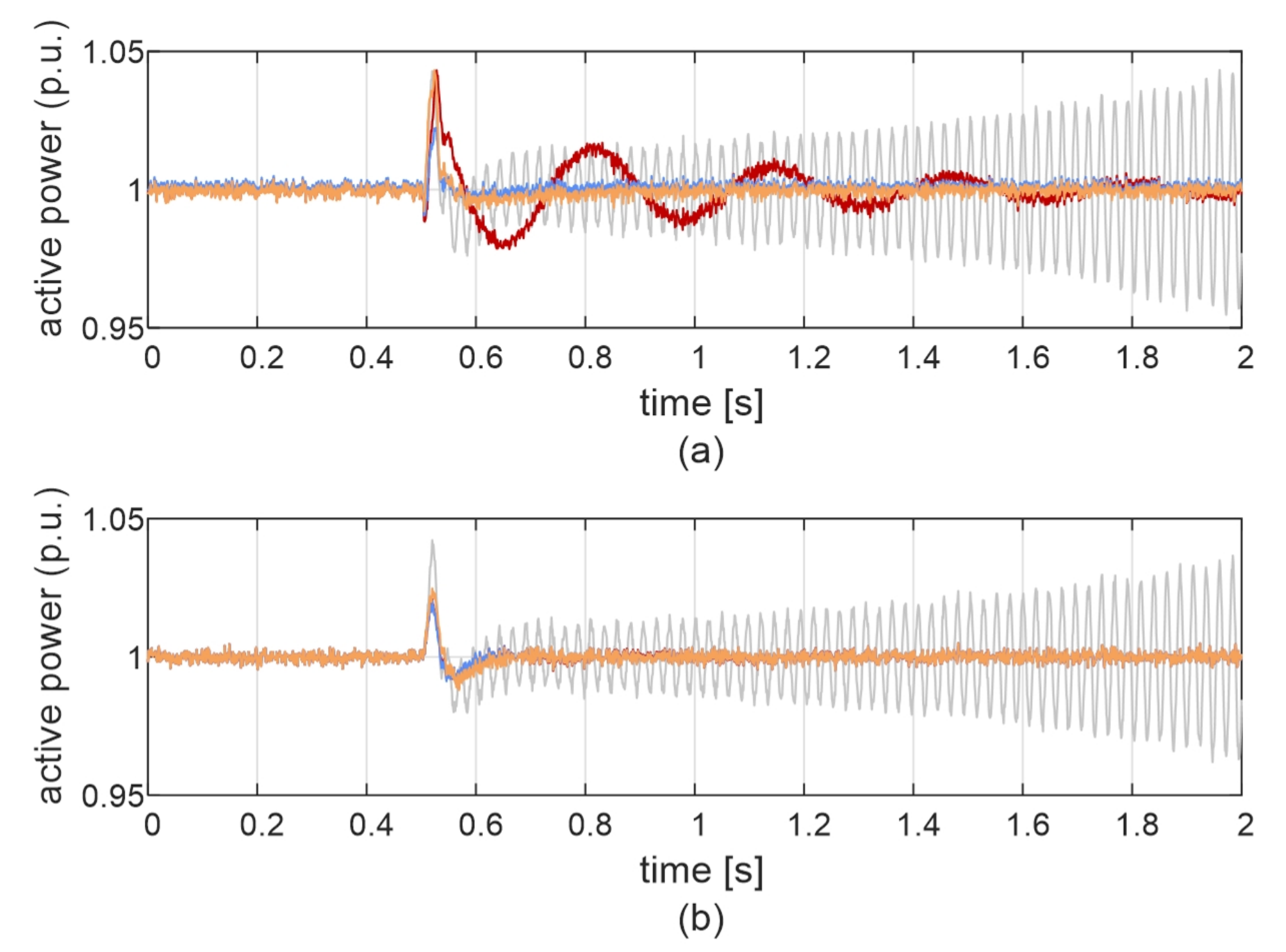}}
	\vspace{-3mm}
        \caption{Time-domain responses of the two-area test system under four cases. At $0.5\,\mathrm{s}$ the infinite bus voltage drops to $0.95\,\mathrm{p.u.}$ for $0.02\,\mathrm{s}$. \textcolor[HTML]{c5c5c5}{\textbf{---}}: Case 1 (All converters employ GFL control);
        \textcolor[HTML]{d95426}{\textbf{---}}: Case 2 (Converter 3 employs GFM control);
        \textcolor[HTML]{5B8FF7}{\textbf{---}}: Case 3 (Converter 3 employs Full DeePConverter (PV));
        \textcolor[HTML]{F7A25B}{\textbf{---}}: Case 4 (Converter 3 employs Full DeePConverter (PQ)).}
        \label{two area result}
	\vspace{-2mm}
\end{figure}

To further validate the performance of DeePConverters in realistic multi-converter systems, we consider a two-area power system with four converters, as illustrated in Fig.~\ref{two area}. In this study, Integral Full DeePConverters operating in both PV and PQ modes are employed. The system parameters and DeePC-related configurations are provided in Appendix~\ref{Appendix A}.

We design four cases (Cases 1$~\sim~$4), where the infinite bus voltage drops by \(0.05 \, \mathrm{p.u.}\) at \(t = 0.5 \, \mathrm{s}\) for $0.02\,\mathrm{s}$ in all cases. Fig.~\ref{two area result} shows the time-domain responses of Converters 3 and 4, as Converters 1 and 2 behave similarly to Converter 4 and are therefore omitted for brevity. In Case 1, all converters operate under PLL-based control, resulting in persistent oscillations due to the interaction between PLLs and the weak grid. When Converter 3 is switched to GFM control in Case 2, the oscillations are significantly suppressed. Further, replacing Converter 3 with Full DeePConverter (PV) in Case 3 or with Full DeePConverter (PQ) in Case 4 also leads to effective suppression of oscillations. Moreover, when DeePConverter control schemes in either PV or PQ mode are employed, the active power fluctuations in Converter 3 after the disturbance are much smaller than those in the GFM case (i.e., Case 2), indicating better dynamic performance. This highlights that DeePConverters in both PV and PQ modes provide strong voltage support for the power networks and stabilize the GFL converters, thus demonstrating both ``self-stabilizing'' and ``stabilizing'' capabilities, with superior performance compared to the conventional methods.

 \begin{figure}[!t]
 \vspace{-4mm}
\centerline{\includegraphics[width=0.75\linewidth]{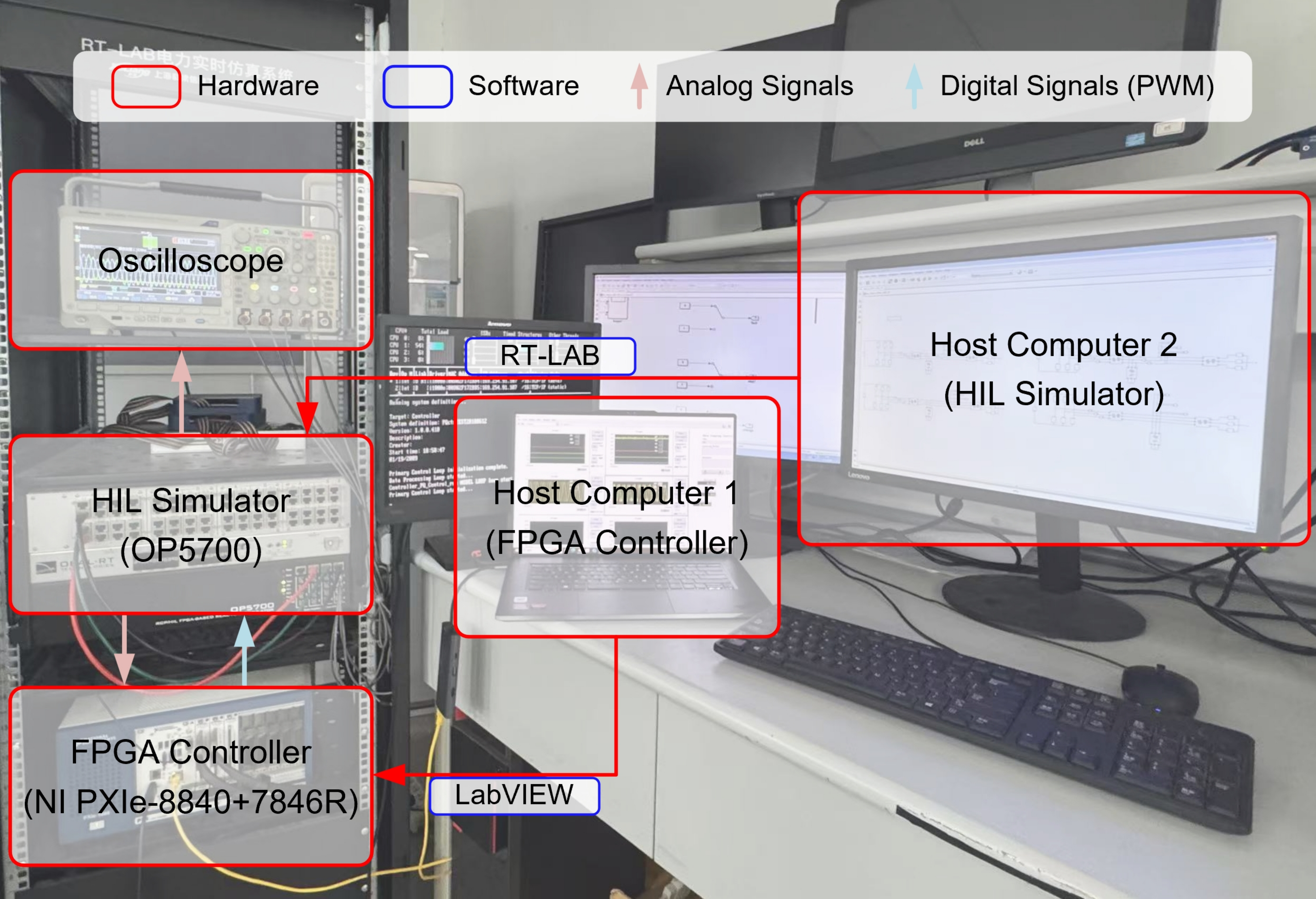}}
	\vspace{-2mm}
        \caption{HIL experimental platform.}
        \label{platform}
	\vspace{-2mm}
\end{figure}

\subsection{HIL Experimental Results}

We now illustrate the effectiveness of the DeePConverters through HIL experiments. The platform, shown in Fig.~\ref{platform}, consists of an OP5700 simulator, an NI PXIe-8840 FPGA controller, two host computers, and an oscilloscope. The FPGA of OP5700 can execute power-stage models with time steps less than \(0.5\,\mu\mathrm{s}\), ensuring high numerical precision. Three single-converter systems (Fig.~\ref{Fig_converter_1}) are emulated: Converter 1 adopts the Integral Power\&Voltage-regulated DeePConverter, Converter 2 a GFL scheme, and Converter 3 a GFM scheme. The power parts of all three systems run on the OP5700 FPGA, while the DeePC algorithm of Converter 1 is executed on the NI FPGA controller using the closed-form solution in~\eqref{eq:mapping} (input/output constraints omitted for speed). The detailed parameters of all three converter systems are provided in Appendix~\ref{Appendix A}.

\begin{figure*}[!t]
 \vspace{-4mm}
\centerline{\includegraphics[width=0.95\linewidth]{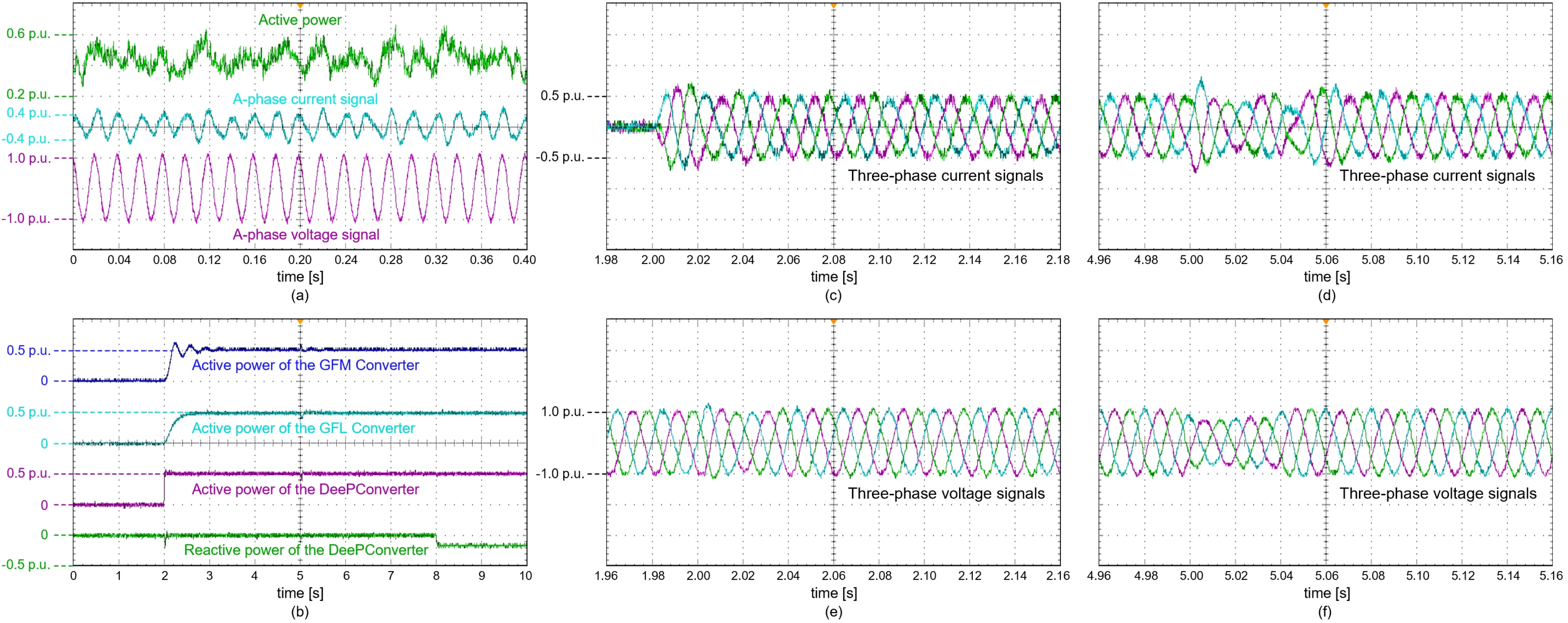}}
	\vspace{-3mm}
        \caption{Time-domain results of the HIL experiment: (a) Responses of Converter 1 during the data-collection period with injected white-noise signals; (b) Responses of the three single-converter systems: Converter 1 (DeePConverter), Converter 2 (GFL Converter), and Converter 3 (GFM Converter). At $2\,\mathrm{s}$ the active power references of all three converters step from $0$ to $0.5\,\mathrm{p.u.}$, at $5\,\mathrm{s}$ the infinite bus voltages drop to $0.75\,\mathrm{p.u.}$ for $0.04\,\mathrm{s}$, and at $8\,\mathrm{s}$ DeePConverter switches from PQ to PV mode; (c) \& (d) Three-phase currents and (e) \& (f) three-phase voltages of DeePConverter at  $2\,\mathrm{s}$ and $5\,\mathrm{s}$, respectively.}
        \label{single_result}
	\vspace{-3mm}
\end{figure*}

The HIL experimental results are summarised in Fig.~\ref{single_result}. During the data-collection phase, band-limited white noise excitation yields persistently exciting trajectories with acceptable power and current variations (Fig.~\ref{single_result} (a)). Fig.~\ref{single_result} (b) shows that the DeePConverter achieves the fastest reference tracking, superior disturbance rejection during the voltage dip, and a smooth, rapid transition from PQ to PV mode, clearly outperforming the GFL and GFM schemes. Waveforms in Fig.~\ref{single_result} (c)-(f) confirm that the three-phase voltages remain tightly regulated and the currents exhibit well-damped transients at \(2\,\mathrm{s}\) and \(5\,\mathrm{s}\). These observations validate the robustness and fast dynamics of the DeePConverter approach.

\section{Conclusions}
This paper presents DeePConverter, a data-driven control architecture that enables grid-connected power converters to perform optimal, robust, and adaptive control without relying on explicit power system models. By leveraging the information hidden in trajectories, DeePConverters achieve data-driven optimal synchronization and power/voltage regulation, which can also replicate and enhance GFL or GFM behavior. A modular design enables flexible configurations of the DeePConverter to replace specific control loops in a plug-in manner, while its integral form enhances steady-state tracking accuracy and its online adaptation mechanism accommodates the time-varying system dynamics. Detailed simulations and HIL tests demonstrate that DeePConverters exhibit excellent tracking, strong disturbance rejection, and both self-stabilizing and stabilizing capabilities. Future work will explore its application across broader power converter types and investigate coordination among multiple DeePConverters.

\appendices
\numberwithin{equation}{section}
\vspace{0mm}
\section{Parameters of the Test Systems}
\label{Appendix A}
See Table~\ref{table:single converter1}, Table~\ref{table:single converter2}, Table~\ref{table:four converter} and Table~\ref{table:HIL test}.
\renewcommand{\arraystretch}{1.25}
\renewcommand{\thetable}{A.\Roman{table}}  
\setcounter{table}{0}  

\begin{table}[h]
	\scriptsize
	\centering
	\vspace{-0mm}
	\caption{Parameters of the Single-Converter System (Subsection~\ref{emulate})}
    \vspace{-1mm}
	\begin{tabular}{|l|l|l|l|}
		\hline
		\multicolumn{4}{|c|}{\bf Base Values for Per-unit Calculation} \\
		\hline
		\multicolumn{1}{|l}{$f_{\rm base} = 50{\rm Hz}$}			
		&\multicolumn{1}{l}{$\omega_{\rm base} = 2\pi f_{\rm base}$}				
        &\multicolumn{1}{l}{$U_{\rm base} = 380{\rm V}$}			
		&\multicolumn{1}{l|}{$S_{\rm base} = 1{\rm kVA}$}								\\
		\hline
		\multicolumn{4}{|c|}{\bf Main Parameters of the Power Part (Per-unit Values)}				\\
		\hline
		\multicolumn{2}{|l}{Converter-side inductors: $L_{\rm F} = 0.05$}			
		&\multicolumn{2}{l|}{Converter-side resistors: $R_{\rm L} = 0.01$}								\\
		\multicolumn{2}{|l}{Grid-side inductors: $L_{\rm g} = 0.1, 0.5$}			
		&\multicolumn{2}{l|}{Grid-side resistors: $R_{\rm g} = 0.02$}									\\
		\multicolumn{2}{|l}{\textit{LCL} capacitors: $C_{\rm F} = 0.05$}		
        
        &\multicolumn{2}{l|}{Active power load: $R_{\rm Load} = 4$}		
										\\
		\hline
		\multicolumn{4}{|c|}{\bf Main Parameters of the GFL Converter (Per-unit Values)}							\\
		\hline
		\multicolumn{2}{|l}{PI gains of the current loop: $0.2, 10$}			
		&\multicolumn{2}{l|}{PI gains of the outer loop: $2, 40$}	\\
            \multicolumn{2}{|l}{PI gains of the PLL: $21, 212$}			
		&\multicolumn{2}{l|}{}	\\	
		\hline
		\multicolumn{4}{|c|}{\bf Main Parameters of the GFM Converter (Per-unit Values)}											\\
		\hline
		\multicolumn{2}{|l}{PI gains of the current loop: $0.5, 20$}			
		&\multicolumn{2}{l|}{PI gains of the voltage loop: $2, 10$}	\\
            \multicolumn{4}{|l|}{Parameters of the swing equation: $J=2$, $D=50$}	\\	
		\hline
		\multicolumn{4}{|c|}{\bf Main Parameters of the DeePConverters (Per-unit Values)}									\\
		\hline
		\multicolumn{4}{|c|}{Common parameters}				\\
		\hline
	\multicolumn{4}{|l|}{$k = 1~~~~~~~~~~T = 700~~~~~~~~~~T_{\text{ini}} = 6~~~~~~~~~~N = 12~~~~~~~~~~ T_\text{S} = 0.001$} \\
		\hline
		\multicolumn{4}{|c|}{Integral Full  DeePConverter 1 (GFL)}		
        \\
		\hline
		\multicolumn{4}{|l|}{$R=I_N \otimes \text{diag}(0.1, 0.1, 0.1)~~~~~~\alpha_1=800\ ~~~~\alpha_2=800\ ~~~~\alpha_3=2000$}			
		\\
        \multicolumn{4}{|l|}{$\lambda_g=10~~~~~~~~~\lambda_{y}=1\times10^4$}
		\\
		\hline
		\multicolumn{4}{|c|}{Integral Full  DeePConverter 2 (GFM)}				  \\
		\hline
		\multicolumn{4}{|l|}{$R=I_N \otimes \text{diag}(0.1, 0.1, 0.1)~~~~~~\alpha_1=100\ ~~~~\alpha_2=1000\ ~~~~\alpha_3=1000$}			
		\\
        \multicolumn{4}{|l|}{$\lambda_g=10~~~~~~~~~\lambda_{y}=1\times10^4~~~~~~~~J=2~~~~~~~~~D=50$}
		\\
		\hline
        \end{tabular}
        \vspace{-3mm}
        \label{table:single converter1}
        \end{table}
        
\noindent
    \begin{minipage}{\linewidth}
        \footnotesize
        \textit{Note:} Integral SPC uses the same parameter settings as DeePConverter~1 (GFL) and DeePConverter~2 (GFM), respectively, except that the DeePC output-slack term and regularization are not used (i.e., $\lambda_y$ and $\lambda_g$ are omitted).
    \end{minipage}

\begin{table}[h]
	\scriptsize
	\centering
	\vspace{-0mm}
	\caption{Parameters of the Single-Converter System \\(Subsection~\ref{transient} and~\ref{adaptive_test})}
    \vspace{-1mm}
	\begin{tabular}{|l|l|l|l|}
		\hline
		\multicolumn{4}{|c|}{\bf Base Values for Per-unit Calculation} \\
		\hline
		\multicolumn{1}{|l}{$f_{\rm base} = 50{\rm Hz}$}			
		&\multicolumn{1}{l}{$\omega_{\rm base} = 2\pi f_{\rm base}$}				
        &\multicolumn{1}{l}{$U_{\rm base} = 380{\rm V}$}			
		&\multicolumn{1}{l|}{$S_{\rm base} = 1{\rm kVA}$}								\\
		\hline
		\multicolumn{4}{|c|}{\bf Main Parameters of the Power Part (Per-unit Values)}				\\
		\hline
		\multicolumn{2}{|l}{Converter-side inductors: $L_{\rm F} = 0.05$}			
		&\multicolumn{2}{l|}{Converter-side resistors: $R_{\rm L} = 0.01$}								\\
		\multicolumn{2}{|l}{Grid-side inductors: $L_{\rm g} = 0.25$}			
		&\multicolumn{2}{l|}{Grid-side resistors: $R_{\rm g} = 0.02$}									\\
		\multicolumn{2}{|l}{\textit{LCL} capacitors: $C_{\rm F} = 0.05$}		
        
        &\multicolumn{2}{l|}{Active power load: $R_{\rm Load} = 4$}		
										\\
		\hline
		\multicolumn{4}{|c|}{\bf Main Parameters of the DeePConverters (Per-unit Values)}									\\
		\hline
		\multicolumn{4}{|c|}{Common parameters}				\\
		\hline
	\multicolumn{4}{|l|}{$k = 1~~~~~~~~~~T = 700~~~~~~~~~~T_{\text{ini}} = 6~~~~~~~~~~N = 12~~~~~~~~~~ T_\text{S} = 0.001$} \\
		\hline
		\multicolumn{4}{|c|}{Integral Self-synchronized DeePConverter PQ}		
        \\
		\hline
		\multicolumn{4}{|l|}{$R=I_N \otimes \text{diag}(0.1, 0.1, 0.1)~~~~~~\alpha_1=800\ ~~~~\alpha_2=800\ ~~~~\alpha_3=2000$}			
		\\
        \multicolumn{4}{|l|}{$\lambda_g=10~~~~~~~~~\lambda_{y}=2.5\times10^4$}
		\\
		\hline
		\multicolumn{4}{|c|}{Integral Self-synchronized DeePConverter PV}				  \\
		\hline
		\multicolumn{4}{|l|}{$R=I_N \otimes \text{diag}(0.1, 0.1, 0.1)~~~~~~\alpha_1=800\ ~~~~\alpha_2=2000\ ~~~\alpha_3=2000$}			
		\\
        \multicolumn{4}{|l|}{$\lambda_g=10~~~~~~~~~\lambda_{y}=2.5\times10^4$}
		\\
		\hline
        \multicolumn{4}{|c|}{Integral Power-regulated DeePConverter PQ}				  \\
		\hline
		\multicolumn{4}{|l|}{$R_{\text{PQ}}=I_N \otimes \text{diag}(0.1, 0.1)~~~~~~~~~Q_{\text{PQ}}=I_N \otimes \text{diag}(0, 0, 2000, 2000)$}			
		\\
        \multicolumn{4}{|l|}{$\lambda_g=1~~~~~~~~~~~\lambda_{y}=1\times10^4$}
		\\
		\hline
	\end{tabular}
    \vspace{-4mm}
	\label{table:single converter2}
\end{table}

\begin{table}[h]
	\scriptsize
	\centering
	\vspace{-0mm}
	\caption{Parameters of the Two-Area Test System}
    \vspace{-1mm}
	\begin{tabular}{|l|l|l|l|}
		\hline
		\multicolumn{4}{|c|}{\bf Base Values for Per-unit Calculation} \\
		\hline
		\multicolumn{1}{|l}{$f_{\rm base} = 50{\rm Hz}$}			
		&\multicolumn{1}{l}{$\omega_{\rm base} = 2\pi f_{\rm base}$}				
        &\multicolumn{1}{l}{$U_{\rm base} = 690{\rm V}$}			
		&\multicolumn{1}{l|}{$S_{\rm base} = 1{\rm MVA}$}								\\
		\hline
		\multicolumn{4}{|c|}{\bf Main Parameters of the Power Part (Per-unit Values)}				\\
		\hline
		\multicolumn{2}{|l}{Converter-side inductors: $L_{\rm F} = 0.05$}			
		&\multicolumn{2}{l|}{\textit{LCL} capacitors: $C_{\rm F} = 0.05$}									\\
		\multicolumn{1}{|l}{\( B_{15} = 4.76 \)}			
		&\multicolumn{1}{l}{\( B_{25} = 6.25 \)}			
		&\multicolumn{1}{l}{\( B_{39} = 4.76 \)}			
		&\multicolumn{1}{l|}{\( B_{49} = 6.25 \)}	\\
		\multicolumn{1}{|l}{\( B_{56} = 66.66 \)}			
		&\multicolumn{1}{l}{\( B_{67} = 66.66 \)}			
		&\multicolumn{1}{l}{\( B_{78} = 28.57 \)}			
		&\multicolumn{1}{l|}{\( B_{710} = 20.83 \)}	\\
		\multicolumn{1}{|l}{\( B_{89} = 50.00 \)}			
		&\multicolumn{1}{l}{\( \tau = 0.10 \)}			
		&\multicolumn{2}{l|}{}	\\
		\hline
		\multicolumn{4}{|c|}{\bf Main Parameters of the GFL Converter (Per-unit Values)}							\\
		\hline
		\multicolumn{2}{|l}{PI gains of the current loop: $0.3, 10$}			
		&\multicolumn{2}{l|}{PI gains of the outer loop: $0.5, 40$}	\\
            \multicolumn{2}{|l}{PI gains of the PLL: $118, 5000$}			
		&\multicolumn{2}{l|}{}	\\	
		\hline
		\multicolumn{4}{|c|}{\bf Main Parameters of the GFM Converter (Per-unit Values)}											\\
		\hline
		\multicolumn{2}{|l}{PI gains of the current loop: $0.3, 10$}			
		&\multicolumn{2}{l|}{PI gains of the voltage loop: $4, 20$}	\\
            \multicolumn{4}{|l|}{Parameters of the swing equation: $J=2$, $D=25$}	\\	
		\hline
		\multicolumn{4}{|c|}{\bf Main Parameters of the DeePConverters (Per-unit Values)}									\\
		\hline
		\multicolumn{4}{|l|}{$k = 1~~~~~~~~~~T = 700~~~~~~~~~~T_{\text{ini}} = 6~~~~~~~~~~N = 12~~~~~~~~~~ T_\text{S} = 0.001$} \\
		\multicolumn{4}{|l|}{$R=I_N \otimes \text{diag}(0.1, 0.1, 0.1)~~~~~~~\alpha_1=200\ ~~~~\alpha_2=200\ ~~~~\alpha_3=200$}			
		\\
        \multicolumn{4}{|l|}{$\lambda_{g}=5~~~~~~~~~\lambda_{y}=2.5\times10^5$}		
		\\
		\hline
	\end{tabular}
    \vspace{-4mm}
	\label{table:four converter}
\end{table}

\begin{table}[t!]
	\scriptsize
	\centering
	\vspace{-0mm}
	\caption{Parameters of the HIL Test System}
    \vspace{-1mm}
	\begin{tabular}{|l|l|l|l|}
		\hline
		\multicolumn{4}{|c|}{\bf Base Values for Per-unit Calculation} \\
		\hline
		\multicolumn{1}{|l}{$f_{\rm base} = 50{\rm Hz}$}			
		&\multicolumn{1}{l}{$\omega_{\rm base} = 2\pi f_{\rm base}$}				
        &\multicolumn{1}{l}{$U_{\rm base} = 690{\rm V}$}			
		&\multicolumn{1}{l|}{$S_{\rm base} = 10{\rm kVA}$}								\\
		\hline
		\multicolumn{4}{|c|}{\bf Main Parameters of the Power Part (Per-unit Values)}				\\
		\hline
		\multicolumn{2}{|l}{Converter-side inductors: $L_{\rm F} = 0.18$}			
		&\multicolumn{2}{l|}{Converter-side resistors: $R_{\rm L} = 0.04$}								\\
		\multicolumn{2}{|l}{Grid-side inductors: $L_{\rm g} = 0.285$}			
		&\multicolumn{2}{l|}{Grid-side resistors: $R_{\rm g} = 0.05$}									\\
		\multicolumn{2}{|l}{\textit{LCL} capacitors: $C_{\rm F} = 0.08$}		
        &\multicolumn{2}{l|}{\textit{LCL} capacitor resistance: $R_{\rm C} = 0.8$}		
										\\
		\hline
		\multicolumn{4}{|c|}{\bf Main Parameters of the GFL Converter (Per-unit Values)}							\\
		\hline
		\multicolumn{2}{|l}{PI gains of the current loop: $0.3, 10$}			
		&\multicolumn{2}{l|}{PI gains of the outer loop: $0.2, 10$}	\\
            \multicolumn{2}{|l}{PI gains of the PLL: $45, 6500$}			
		&\multicolumn{2}{l|}{}	\\	
		\hline
		\multicolumn{4}{|c|}{\bf Main Parameters of the GFM Converter (Per-unit Values)}											\\
		\hline
		\multicolumn{2}{|l}{PI gains of the current loop: $0.8, 12$}			
		&\multicolumn{2}{l|}{PI gains of the voltage loop: $6, 20$}	\\
            \multicolumn{4}{|l|}{Parameters of the swing equation: $J=3$, $D=20$}	\\	
		\hline
		\multicolumn{4}{|c|}{\bf Main Parameters of the DeePConverter (Per-unit Values)}									\\
		\hline
		\multicolumn{4}{|l|}{$k = 1~~~~~~~~~~T = 700~~~~~~~~~~T_{\text{ini}} = 6~~~~~~~~~~N = 12~~~~~~~~~~ T_\text{S} = 0.001$} \\
		\multicolumn{4}{|l|}{$R_{\text{PV}}=I_N \otimes \text{diag}(0.1, 0.1)~~~~~~~~~Q_{\text{PV}}=I_N \otimes \text{diag}(500, 0, 500, 0)$}			
		\\
		\multicolumn{4}{|l|}{$R_{\text{PQ}}=I_N \otimes \text{diag}(0.1, 0.1)~~~~~~~~~Q_{\text{PQ}}=I_N \otimes \text{diag}(0, 0, 500, 500)$}		
		\\
        \multicolumn{4}{|l|}{$\lambda_{g\text{PV}}=100~~~~~~~\lambda_{g\text{PQ}}=100~~~~~~\lambda_{y\text{PV}}=1.0\times10^4~~~~~\lambda_{y\text{PQ}}=1.0\times10^4$}		
		\\
		\hline
	\end{tabular}
    \vspace{-3mm}
	\label{table:HIL test}
\end{table}

    \normalem
    \bibliographystyle{IEEEtran}
	\bibliography{references}

\end{document}